\documentclass{aa}

\usepackage{graphicx}
\usepackage{txfonts}
\usepackage{comment}
\usepackage[utf8]{inputenc}
\usepackage[T1]{fontenc}
\usepackage[english]{babel}
\usepackage{natbib}

\usepackage{url}
\usepackage{color}

\usepackage{amsmath}

\definecolor{azg}{rgb}{0.0, 0.46, 0.37}
\definecolor{al}{rgb}{0.90,0.35,0}
\definecolor{grey}{rgb}{0.5,0.5,0.5}
\usepackage{ulem}

\begin{document}

  \title{Kernel-phase Detection Limits : Hypothesis Testing\\ and the Example of JWST NIRISS Full Pupil Images}
  \titlerunning{Kernel Detection Limits}

  \authorrunning{Ceau, Mary, Greenbaum}
  \author{A. Ceau \inst{1}, D. Mary \inst{1}, A. Greenbaum \inst{2}, F. Martinache \inst{1}, A. Sivaramakrishnan \inst{3}, R. Laugier\inst{1}, \\M. N'Diaye \inst{1}
     }
     \institute{Université Côte d'Azur, Observatoire de la Côte d'Azur, CNRS, Laboratoire Lagrange, France
     \and Department of Astronomy, University of Michigan, Ann Arbor, MI 48109, USA
     \and Space Telescope Science Institute, Baltimore, MD 21218, USA
     }
  \date{Received June 30th 2018}

 
 \abstract
{The James Webb Space Telescope will offer high-angular resolution observing capability in the near-infrared with masking interferometry on NIRISS, and coronagraphic imaging on NIRCam \& MIRI. Full aperture kernel-phase based interferometry complements these observing modes by allowing to probe for companions at small angular resolution while preserving the telescope throughput.}
{Our goal is to  derive both theoretical and operational  contrast detection limits for the kernel-phase analysis of JWST NIRISS full-pupil observations by using tools from hypothesis testing theory. The study is immediately applied to observations of faint brown dwarfs with this instrument, but the tools and methods introduced here are applicable in a wide variety of contexts.}
{We construct a statistically independent set of observable quantities from a collection of aberration-robust kernel phases. Three  detection tests based on these observable quantities are designed and analysed, all having the property of guaranteeing a constant false alarm rate for phase aberrations smaller than about one radian. One of these tests, the Likelihood Ratio or Neyman-Pearson test, provides a theoretical performance bound  for any detection test.}
{The operational detection method considered here is shown to exhibit only marginal power loss with respect to the theoretical bound. In principle, for the test set to a false alarm probability of $1\%$, companion  at contrasts reaching $10^3$ at separations of $200$ mas around objects of magnitude $14.1$ are detectable with probability $68\%$. For the brightest objects observable using the full pupil of JWST and NIRISS, contrasts of up to $10^4$ at separations of $200$ mas could be ultimately achieved, barring significant wavefront drift. We also provide a statistical analysis of the uncertainties affecting the  contrasts and separations that are estimated for the detected companions.}
{The proposed detection method is close to the ultimate bound and offers guarantees over the probability of making a false detection for binaries, as well as over the error bars for the estimated parameters of the binaries that will be detected by JWST NIRISS. This method is not only applicable to JWST NIRISS but to any imaging system with adequate sampling.}
\maketitle
%

\section{Introduction}
In the past few years, many nearby brown dwarfs have been discovered thanks to the Wide-field Infrared Survey Explorer (WISE) sky survey \citep{Wright2010ThePerformance, Cushing2011THE,Schneider2015HUBBLEEXPLORER}. These newly discovered objects present an observational challenge due to their intrinsically low luminosities. 
Some of these brown dwarfs have been observed by the Hubble Space Telescope (HST), mostly for proper motion and parallax measurements \citep[e.g.][]{Marsh2013ParallaxesT}. While previous studies have searched for companions, they lack the sensitivity in the optical and the near infrared to achieve high enough contrasts to detect very low mass companions \citep[e.g.][]{Fontanive2018ConstrainingType}. 
High-angular resolution observations are also possible from the ground using either adaptive optics or optical interferometry. Cool dwarfs are however intrinsically faint objects and therefore fall short of the requirements of either technique, unless it is assisted by laser guide stars \citep{Bernat2010AOPTICS}.

Issues limiting the quality of ground based observations, such as sky background or atmospheric perturbations can be alleviated by observing from space. When launched, the James Webb Space Telescope \citep[JWST,][]{Gardner2006TheTelescope} will be the largest ever space telescope, and provide unparalleled sensitivity for studying faint, cool dwarfs. With a 6.5-meter primary mirror, and an instrument suite covering the 0.6-25.5 $\mu$m wavelength range, the theoretical angular resolution of this telescope respectively ranges from 20 to 800 mas. For a nearby object located less than 20 pc away, this translates in the ability to resolve structures present within a few astronomical units (AU) of the central source.

However, even for instruments capable of very high angular resolution, the glare from an object can drown out the light of faint surrounding structures. This issue is usually addressed by using coronagraphy and JWST's instrumentation offers several coronagraphs inside NIRCam and MIRI, with inner working angles ranging from 300 to 800 mas.
To probe the innermost parts of nearby systems, inside the inner working angles of the coronagraphs, interferometry offers a viable alternative. In that scope, on board JWST, NIRISS offers the aperture masking interferometer (AMI) observing mode \citep{Sivaramakrishnan2012Non-redundantJWST-NIRISS} with a non-redundant mask (NRM) located in the instrument pupil wheel. AMI enables the detection of objects with lower contrasts, but at narrower separations compared to what can be achieved by JWST's coronagraphs. AMI is expected to have sufficient performance to address yet unanswered questions in the fields of active galaxy nuclei \citep{Ford2014ActiveTelescope}, planetary formation, exoplanet \citep{Artigau2014NIRISSOpportunities}, as well as follow-ups on astrometry measurements from the GAIA mission, or on ground based extreme adaptive optics (AO) surveys. In the case of binary point sources in non-coronagraphic modes, contrast ratios as high as 10 magnitudes ($10^4$) for the brightest companions at 130 mas can be attained using AMI \citep{Sivaramakrishnan2012Non-redundantJWST-NIRISS,Greenbaum2015ANDATA,Greenbaum2018GPIModeling}.

AMI achieves its best performance by taking advantage of self-calibrating observable quantities called closure-phases \citep{Jennison1958AExtent}. This technique, first developped for radio interferometry and later ported to the optical regime \citep{Baldwin1986ClosureImaging} was adapted to single dish telescopes using a non-redundant aperture mask. Initially used in seeing-limited observing conditions \citep{Nakajima1989Diffraction-limitedStars}, the technique eventually took advantage of the development of AO \citep{Tuthill2006PinwheelsCluster} allowing for stabilized longer exposure modes and the ability to observe fainter objects. NRM interferometry is now routinely used and has led to a variety of studies, e.g.  \citet{Sallum2015AccretingDisk,Kraus2008MappingScorpius,Kraus2011MappingTaurus-Auriga}.


Kernel-phase generalises the idea of closure-phase to apertures of arbitrary shapes, and can reliably be used when aberrations are smaller than about one radian \citep{Martinache2010Kernel-PhaseInteferometry}. This method can therefore be used on images acquired using any instrument onboard JWST, provided that the instrument pupil geometry is  accurately modelled. It is therefore useable on full-pupil images as well as on AMI/NRM closure phases. The Kernel method has already been used successfully to uncover new brown dwarf binaries with HST observations, as reported by \citet{Pope2013DancingInterferometry}. Full-aperture kernel-phase and AMI closure-phase cover the same parameter space but with its lower throughput ($\sim$15 \%), AMI is suited for the observation of bright targets that would otherwise saturate the instrument,as well as for observations where aberrations are too important to fall into the linear regime covered by the kernel method.

Kernel- and closure-phase rely on exploiting the phase of the Fourier transform (also referred to as the complex visibility) of the image. The image must satisfy the Nyquist-sampling requirement (platescale smaller than $0.5\;\lambda/D$), although small grid dithering allows observers to reconstruct a Nyquist Sampled image for other filters.  Saturation should be avoided, although recovery is still possible. \citep{Laugier2019Recovering494AB}. For a filter to be fully exploitable, its shortest wavelength must respect the sampling criterion. For the $6.5\ \mathrm{m}$ diameter of the primary mirror of JWST, this means that the filters compatible with a Kernel-phase analysis are:

\begin{itemize}
\item{for NIRCam in the short wavelength channel ($0.6-2.3 \mu$m), with a platescale of 31 mas/pixel: F212N}
\item{for NIRCam in the long wavelength channel ($2.4 - 5.0 \mu$m), with a platescale of 63 mas/pixel: F430M, F460M, F466N, F470N and F480M.}
\item{for NIRISS, with a platescale of 65 mas/pixel \citep{STSCI2018NIRISSImaging}: F430M and F480M.}
\item{for MIRI, with a platescale of 110 mas/pixel: all filters but F770W and F780W.}
\end{itemize}

Kernel detection limits for NIRCam have been computed by \citet{Sallum2019ComparingCharacterization}, for the F430M and F480M filters, as well a for NIRISS AMI in those same bands. The present work aims at setting  a general statistical framework for the theoretical and operational detection limits of the Kernel approach, with focus on guarantees for the actual false alarm rate of the implemented detection method. As for the practical results, we investigate various aspects of the detection limits achievable for full-aperture NIRISS observations in the F480M filter.

Section \ref{sec:ker_stats} will remind how kernel-phases are constructed, present the corresponding statistical model and introduce three statistical tests that will later be used to determine contrast detection limits. Section \ref{sec:results} will show how the method is applied to simulated images by JWST NIRISS. For several objects representative of the Y dwarfs discovered by WISE, this part highlights the need for estimating the noise covariance matrix, compares the performance of proposed detection tests and analyses the statistical uncertainty resulting in the estimation of the parameters of the detected binaries.
For the remainder of this paper, an italicised lowercase letter such as $a$ will denote a real or complex number, a bold lowercase italicised letter such as $\boldsymbol{a}$ will denote a vector, a bold italicised uppercase letter such as $\boldsymbol{A}$ will denote a matrix, and a hat such as $\widehat{b}$ will denote the Maximum Likelihood Estimate, or MLE, of an unknown parameter $b$.

\section{Kernel approach and statistical models}
\label{sec:ker_stats}
\begin{figure}
\includegraphics[width=6cm]{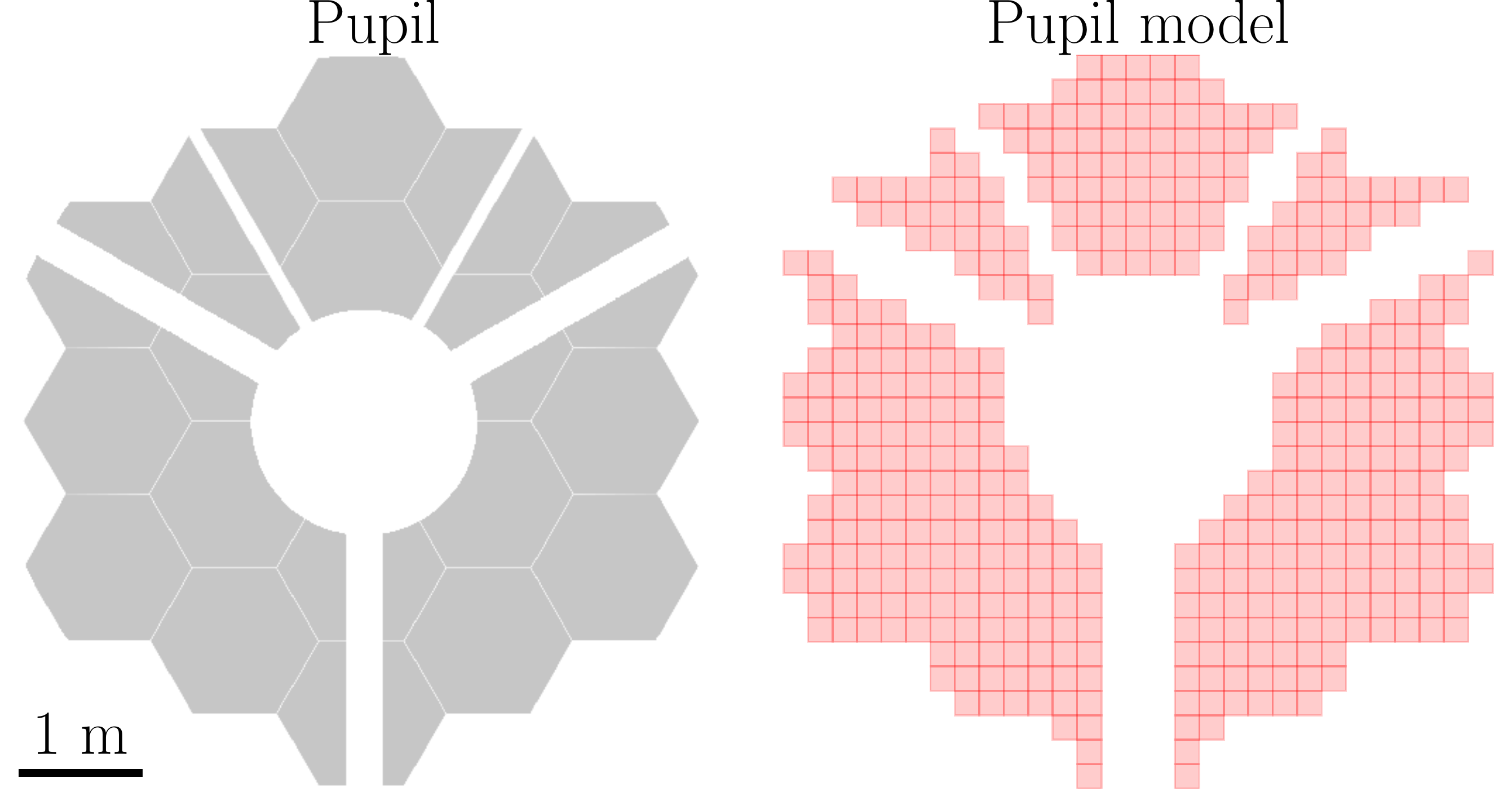}~
\includegraphics[width=3cm]{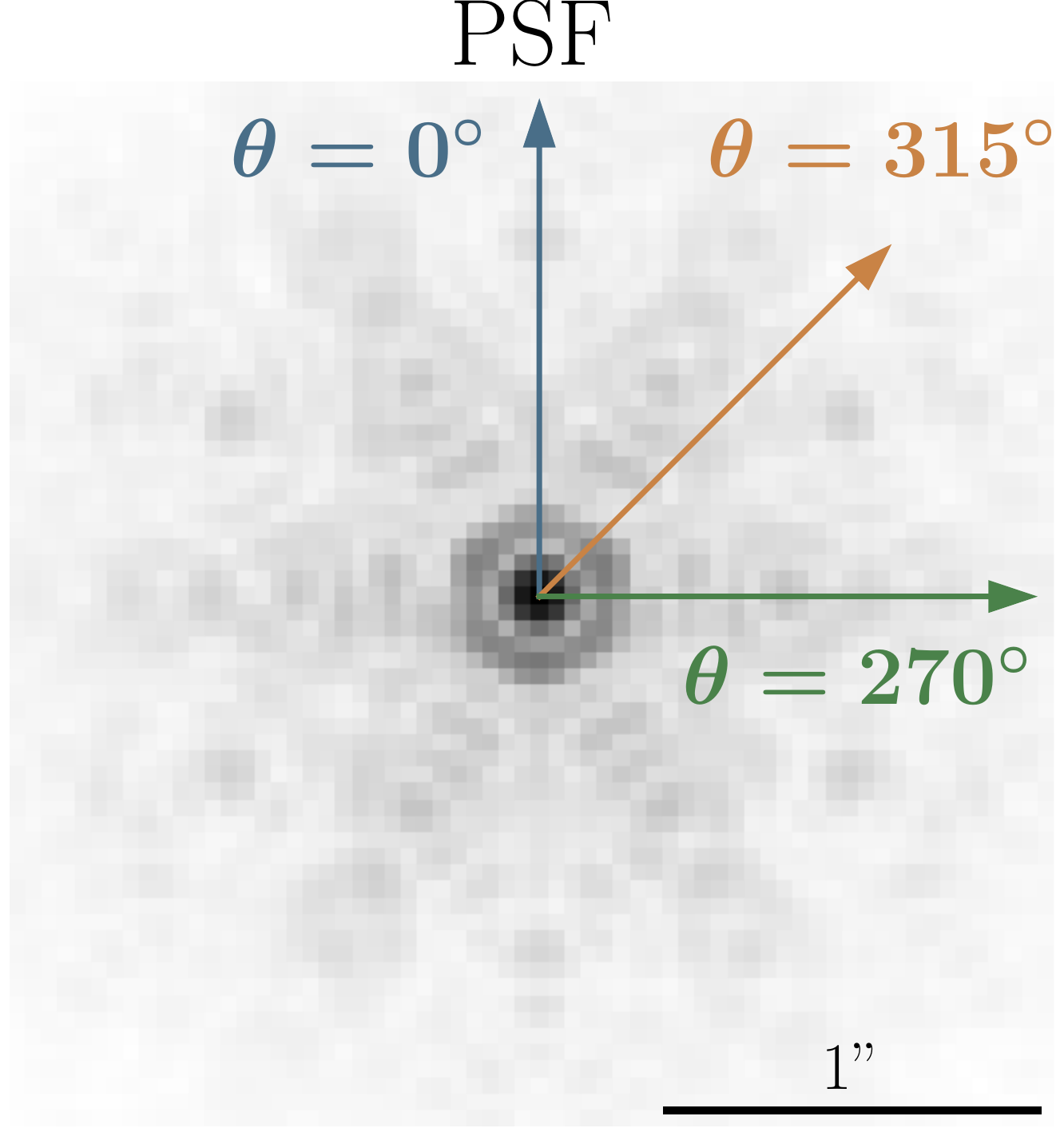}
\caption{Left: the entrance pupil for JWST. Centre: a discrete model of the pupil. The pupil is modeled by an array of subpupils, enabling  the use of the kernel method. Right: The simulated PSF for NIRISS using the 480M filter, represented using a non-linear colour scale. The coloured arrows represent the directions along which the simulated companions are placed.}
\label{fig:pupil_model}
\end{figure}
\subsection{The Kernel Approach}
\label{sec:kernel}
The kernel framework introduced by
\citet{Martinache2010Kernel-PhaseInteferometry} describes 
diffraction-dominated images produced by the mostly-continuous aperture of a
telescope as if they were the interference pattern formed by a discrete array of virtual sub-apertures laid out on a regular grid of finite step. Although any pupil model can in principle yield kernel-phases, using a regularly spaced grid  allows to encode simply and effectively the redundancy of the filled aperture. The fidelity of the discrete representation of the
continuous aperture increases with the density the grid. In practice however, the size of the grid step ($s$) translates into a cut-off frequency $\lambda/s$ that is matched to the field of view over which the diffractive signal is recorded. The example of the discrete representation of the JWST entrance aperture along with an image of the theoretical point spread function (PSF) of the original aperture are shown in Fig. \ref{fig:pupil_model}. The entrance pupil is a combination of the entrance pupil and of an additional pupil-plane mask, CLEARP.

Kernel-phases are formed from a linear combination of the phase measured in the Fourier transform of the image. For a given wavelength, the discrete grid describing the original aperture also defines the sampling of the Fourier space via the coordinates and redundancies of the different baselines. The values of the Fourier transform of an image for the selected
spatial frequencies are collected in the complex visibility vector $\boldsymbol{\bf v}$. The phase vector $\boldsymbol{\phi}$ is defined as the argument of the complex visibility
\begin{equation}
\boldsymbol{\phi}:=\angle{\boldsymbol{{\bf{v}}}}.
\end{equation}
In the optical path of a diffraction-limited instrument,  unknown and potentially evolving aberrations result in a variable PSF that degrade the image quality. According to \citet{Martinache2010Kernel-PhaseInteferometry}, in the small
aberration regime and for simple (i.e. non-coronagraphic) images, a linear model relates the phase $\boldsymbol{\phi}$ measured in the Fourier space  to the true phase of the observed object $\boldsymbol{\phi_0}$ and to the aberration  phase $\boldsymbol{\varphi}$ present across the aperture:
\begin{equation} 
\boldsymbol{\phi}=\boldsymbol{\phi_0}+\boldsymbol{A\varphi},
\label{eq:Aphi}
\end{equation}
\noindent where $\boldsymbol{A}$ is a phase transfer matrix, encoding how the aberrations in each subaperture will propagate to the Fourier phase of the image. Its properties depend on the discrete representation of the aperture. The discrete model of JWST's aperture featured in Fig. \ref{fig:pupil_model} is made of $m=452$ virtual sub-apertures, placed on a grid with a step size of 20 cm that form $n = 1363$ distinct baselines, resulting in a full rank phase transfer matrix $\boldsymbol{A}$ of dimensions $1363 \times 452$. The kernel matrix $\boldsymbol{K}$ is defined as a $p \times n$ matrix that verifies
\begin{equation} 
\boldsymbol{KA}=\boldsymbol{0}.
\end{equation}
The kernel matrix cancels phase perturbation to the first order \citep{Ireland2013PhaseMasking}. With the chosen model, this matrix makes it possible to form a vector of kernel-phases $\boldsymbol{k}$ of size $(p\times1)$, with $p = 887$,  defined as:
\begin{equation} 
\boldsymbol{k}:=\boldsymbol{K\phi}.
\label{eq:ker_phi_def}
\end{equation}
 \noindent The kernel matrix $\boldsymbol{K}$ represents the left-nullspace of the transfer matrix $\boldsymbol{A}$, and is computed from its singular value decomposition. The discrete representation of the aperture, the associated phase transfer matrix $\boldsymbol{A}$ and the kernel matrix $\boldsymbol{K}$ can be generated using a specially designed Python package called \texttt{XARA}\footnote{XARA is available at http://github.com/fmartinache/xara/}, that also offers the basic tools to extract kernel-phases from images.

\subsection{Statistical modelling and hypothesis tests}
\label{sec:stat_models}

Given a data image, how likely is it that a companion is present? The present study proposes to tackle this question through statistical hypothesis testing. A hypothesis test compares a test statistic (noted $T$) to a threshold ($\xi$), and has the general form

\begin{equation}
T(\boldsymbol{y})\underset{\mathcal{H}_0}{\overset{\mathcal{H}_1}{\gtrless}} \xi,
\label{gentest}
\end{equation}

\noindent
where $\boldsymbol{y}$ is the data under test (obtained from the image) and the test statistic $T(\boldsymbol{y})$ is a real random variable.  In \eqref{gentest}, the null hypothesis $\mathcal{H}_0$ (noise only) is claimed if $T(\boldsymbol{y})<\xi$ and the alternative hypothesis $\mathcal{H}_1$ (noise + companion)  is claimed otherwise.  If the distribution of $T$ can be known, the probability of false alarm can be controlled by the value of the test threshold $\xi$. \\
The performance of a detection test are given by its probability of false alarm ($P_{FA}$, the probability that a detection occurs under $\mathcal{H}_0$) and its probability of detection ($P_{DET}$, the probability that a detection occurs under $\mathcal{H}_1$):
\begin{equation}
\begin{split}
P_{FA}&:=\mathrm{Pr}\big(T(y)>\xi\ ;\ \mathcal{H}_0\big),\\
P_{DET}&:=\mathrm{Pr}\big(T(y)>\xi\ ;\ \mathcal{H}_1\big) \ .
\label{eq:pfa_pdet}
\end{split}
\end{equation}

The \textit{power} of a test is its $P_{DET}$ at a given $P_{FA}$: the higher the $P_{DET}$ for a given $P_{FA}$, the more powerful the test. It can be conveniently represented as a receiver operating characteristic (ROC) curve, $P_{DET}$ as a function of  $P_{FA}$.

Turning back to our detection problem,
in the absence of noise the ker phases can take the values 
\begin{equation}
\begin{cases}
\boldsymbol{k}=\boldsymbol{0},     &\textnormal{if the target is centrosymmetric or}\\
\boldsymbol{k}=\boldsymbol{K\phi_0},  &\textnormal{if the target presents asymmetries.} 
\end{cases}
\label{eq:noiseless_hyp}
\end{equation}

The noises affecting the images, propagate into the Fourier phases and consequently into the ker-phases. As we shall see in the next Section, the noise on the kernels can be modelled by a correlated Gaussian distribution with a covariance noted $\boldsymbol{\Sigma}$. If this  matrix is known, we can construct  a vector $\boldsymbol{y}$ of `whitened'  kernel-phases which are decorrelated (hence independent), and similarly  a vector $\boldsymbol{x}$ of whitened theoretical kernel-phases corresponding to the signature of the target:

\begin{eqnarray}
\boldsymbol{y}&:=& \boldsymbol{\Sigma}^{-\frac{1}{2}}\boldsymbol{k},\\
\boldsymbol{x}&:= &\boldsymbol{\Sigma}^{-\frac{1}{2}}\boldsymbol{K\phi_0}.
\end{eqnarray}
This leads to the following statistical hypotheses:
\begin{equation}
\begin{cases}
\mathcal{H}_0: \boldsymbol{y}=\boldsymbol{\epsilon}\\
\mathcal{H}_1: \boldsymbol{y}=\boldsymbol{x}+\boldsymbol{\epsilon} 
\end{cases},\quad \boldsymbol{\epsilon}\sim \mathcal{N}(\boldsymbol{0}, \boldsymbol{I}),
\label{eq:H0H1xy}
\end{equation}
where $\boldsymbol{\epsilon}$ is a $p\times 1$ noise vector with independent and identically distributed Gaussian entries ( thanks to the whitening), and $\mathcal{N}(\boldsymbol{0},\boldsymbol{I})$ denotes the standard normal distribution (the covariance of $\boldsymbol{\epsilon}$ is the Identity matrix, $\boldsymbol{I}$).

\subsubsection{Known signature in white Gaussian noise}
For the problem defined in \eqref{eq:H0H1xy}, the most powerful test is the {likelihood  ratio (LR)}, or Neyman-Pearson (NP) test \citep{Neyman1933OnHypotheses}. For this test,
the companion signature $\boldsymbol{x}$ must be known.  The NP test is defined as 
\begin{equation}
\frac{\ell(\boldsymbol{x};\boldsymbol{y})}{\ell(\boldsymbol{0};\boldsymbol{y})}\underset{\mathcal{H}_0}{\overset{\mathcal{H}_1}{\gtrless}} \eta\ ,
\label{eq:LR}
\end{equation}

\noindent
where $\ell(\boldsymbol{x};\boldsymbol{y})$ is the likelihood of the signature $\boldsymbol{x}$ given the data $\boldsymbol{y}$ and $\eta$ an adjustable threshold. For the Gaussian, white noise considered here, the likelihood is \citep{Scharf1994MatchedDetectors}:

\begin{equation}
\ell(\boldsymbol{x};\boldsymbol{y})=(2\pi)^{\displaystyle{-\frac{p}{2}}}\mathrm{exp}\left({\displaystyle{-\frac{1}{2}(\boldsymbol{x}-\boldsymbol{y})^T(\boldsymbol{x}-\boldsymbol{y})}}\right) \ ,
\label{eq:def_likelihood}
\end{equation}

\noindent
$p$ being the length of the kernel phase vector. Similarly, for $\mathcal{H}_0$ with $\boldsymbol{x}=\bf{0}$. The LR test in Eq. \eqref{eq:LR} becomes
\begin{equation}
\mathrm{exp}\left({\displaystyle{-\frac{1}{2}(\boldsymbol{x}^T\boldsymbol{x}-2\boldsymbol{y}^T\boldsymbol{x})}} \right)\underset{\mathcal{H}_0}{\overset{\mathcal{H}_1}{\gtrless}} \eta \ .
\label{eq:LR2}
\end{equation}

Taking the logarithm of Eq. \eqref{eq:LR2} and noting $\xi:=\eta+\frac{1}{2}\boldsymbol{x}^T\boldsymbol{x}$ leads to the test 
\begin{equation}
T_{\mathrm{NP}}(\boldsymbol{y},\boldsymbol{x})=\boldsymbol{y}^T\boldsymbol{x} \underset{\mathcal{H}_0}{\overset{\mathcal{H}_1}{\gtrless}} \xi.
\label{eq:NP}
\end{equation}
Hence, the NP test amounts to comparing the dot product of the data with the signature to a threshold. 
The distribution of $T_{NP}$  can be analytically determined under $\mathcal{H}_0$ and $\mathcal{H}_1$:
\begin{equation}
\begin{cases}
\mathcal{H}_0: T_{\mathrm{NP}}(\boldsymbol{y})\sim \mathcal{N}(0 ,\boldsymbol{x}^T\boldsymbol{x}\big), \\
\mathcal{H}_1: T_{\mathrm{NP}}(\boldsymbol{y},\boldsymbol{x})\sim \mathcal{N}\big(\boldsymbol{x}^T\boldsymbol{x},\boldsymbol{x}^T\boldsymbol{x}\big) \ .
\end{cases}
\label{eq:NP_distributions}
\end{equation}
Denoting by $\mathcal{N}(0,1)$ a standard normal variable and by $\mathcal{F}_{\mathcal{N}}$ its cumulative distribution function (CDF), Def. \eqref{eq:pfa_pdet} and Eq. \eqref{eq:NP_distributions}, the $P_{FA}$ and $P_{DET}$ can be derived as
\begin{equation}
\begin{cases}
P_{FA}^{T_{\mathrm{NP}}}(\xi) = 1-\mathcal{F}_{\mathcal{N}}\Big(\displaystyle{\frac{\xi}{\sqrt{\boldsymbol{x}^T\boldsymbol{x}}}}\Big), \\
P_{DET}^{T_{\mathrm{NP}}}(\xi) = 1-\mathcal{F}_{\mathcal{N}}\Big(\displaystyle{\frac{\xi-\boldsymbol{x}^T\boldsymbol{x}}{\sqrt{\boldsymbol{x}^T\boldsymbol{x}}}}\Big), \\
\end{cases}
\label{pfanp}
\end{equation}
which, for the purpose of plotting ROC curves, combine to 

\begin{equation}
P_{DET}^{T_{\mathrm{NP}}}(P_{FA}^{T_{\mathrm{NP}}}) = 1-\mathcal{F}_{\mathcal{N}}\left(\mathcal{F}_{\mathcal{N}}^{-1}(1-P_{FA}^{T_{NP}})-\sqrt{\boldsymbol{x}^T\boldsymbol{x}}\right)\ .\\
\label{eq:NP_PfaPdet}
\end{equation}

This test is the most powerful for the considered model, and will serve as the benchmark against which any other detection tests can be evaluated.

Implementing the NP test (Eq. \eqref{eq:NP}) requires to know the target signature $\boldsymbol{x}$ (namely, contrast and {position} if $\boldsymbol{x}$ correspond to a companion). In practical situations however, $\boldsymbol{x}$ is often partially or even fully unknown. This leads to consider the statistical model
\begin{equation}
\begin{cases}
\mathcal{H}_0: \boldsymbol{y}=\boldsymbol{\epsilon}, \\
\mathcal{H}_1: \boldsymbol{y}=\boldsymbol{x} + \boldsymbol{\epsilon},\quad \boldsymbol{x} \in \mathcal{X}
\end{cases}
\label{eq:H0H1NP}
\end{equation}
where $\mathcal{X}$ is a space describing some prior information about $\boldsymbol{x}$. We will consider below two cases: completely unknown signature ($\mathcal{X} =\mathbb{R}^p$) and signature of a binary with unknown contrast and separation ($\mathcal{X}$ is then the space spanned by all possible binary signatures). A classical approach when some parameters describing the target $\boldsymbol{x}$ are unknown is to inject its Maximum Likelihood {{estimate}} (MLE, noted  $\boldsymbol{\widehat{x}}$) in place of $\boldsymbol{{x}}$ in the likelihood ratio \eqref{eq:LR}. The MLE is defined by
\begin{equation}
\widehat{\boldsymbol{x}}:=\underset{\boldsymbol{z}\in \mathcal{X}}{\mathrm{argmax}}\ \ell(\boldsymbol{z};\boldsymbol{y})
\end{equation}
and injecting the MLE in the LR  leads to the so-called generalised likelihood ratio (GLR) defined as
\begin{equation}
\frac{\underset{\boldsymbol{z}\in \mathcal{X}}{\mathrm{max}}\ \ell(\boldsymbol{z};\boldsymbol{y})}{\ell(\boldsymbol{0};\boldsymbol{y})} \underset{\mathcal{H}_0}{\overset{\mathcal{H}_1}{\gtrless}} \eta \ 
\quad
\Leftrightarrow \quad  \frac{ \ell(\boldsymbol{\widehat{x}};\boldsymbol{y})}{\ell(\boldsymbol{0};\boldsymbol{y})} \underset{\mathcal{H}_0}{\overset{\mathcal{H}_1}{\gtrless}} \eta.
\label{eq:GLR}
\end{equation}

\subsubsection{Completely unknown $\boldsymbol{x}$ signature}
If we assume as a worst case situation that nothing is known about the signature $\boldsymbol{x}$, we have $\mathcal{X}=\mathbb{R}^p$. The likelihood in \eqref{eq:def_likelihood} is maximised for $\widehat{\boldsymbol{x}}=\boldsymbol{y}$, and injecting this value in Eq. \eqref{eq:GLR} yields
\begin{equation}
\frac{\mathrm{exp}\left({\displaystyle{-\frac{1}{2}(\boldsymbol{y}-\boldsymbol{y})^T(\boldsymbol{y}-\boldsymbol{y})}}\right)}{\mathrm{exp}\left({\displaystyle{-\frac{1}{2}(\boldsymbol{y}-\boldsymbol{0})^T(\boldsymbol{y}-\boldsymbol{0})}}\right)} \underset{\mathcal{H}_0}{\overset{\mathcal{H}_1}{\gtrless}} \xi' \,
\label{eq:ED1}
\end{equation}
with $\xi '$ a threshold. Taking the logarithm of Eq. \eqref{eq:ED1}, we obtain the test: 
\begin{equation}
T_{\mathrm{E}}(\boldsymbol{y}):={\|}\boldsymbol{y}\|^2\underset{\mathcal{H}_0}{\overset{\mathcal{H}_1}{\gtrless}} \xi \ .
\label{eq:ED2}
\end{equation}
This test uses the measured  squared norm of the signal as a test statistic and is called an energy detector (hence $T_{\mathrm{E}}$). Its statistic is distributed as:
\begin{equation}
\begin{cases}
\mathcal{H}_0: T_{\mathrm{E}}(\boldsymbol{y})\sim \chi^2_p(\lambda^2=0), \\
\mathcal{H}_1: T_{\mathrm{E}}(\boldsymbol{y})\sim \chi^2_p(\lambda^2=\boldsymbol{x}^T\boldsymbol{x}) \ .
\end{cases}
\label{eq:ED_distributions}
\end{equation}
Denoting by $\mathcal{F}_{\chi^2_p(\lambda^2)}$ the CDF of a $\chi^2_p(\lambda^2)$ random variable with $p$ degrees of freedom and non-centrality parameter $\lambda$, we obtain:
\begin{equation}
\begin{cases}
P_{FA}^{T_{\mathrm{E}}}(\xi) = 1-\mathcal{F}_{\chi^2_p(0)}(\xi), \\
P_{DET}^{T_{\mathrm{E}}}( \xi) = 1-\mathcal{F}_{\chi^2_p(\boldsymbol{x}^T\boldsymbol{x})}(\xi). \\
\end{cases}
\label{eq:ED_PfaPdet}
\end{equation}
{We note that test $T_E$ in Eq. \eqref{eq:ED2} was previously used in the literature, e.g.  in \citet{Zwieback2016ARegions} and \citet{LeBouquin2012OnInterferometry} (although not identified as a GLR), with the
$P_{FA}$ reported in Eq. \eqref{eq:ED_PfaPdet}.} 

The expressions above combine into:
\begin{equation}
P_{DET}^{T_{\mathrm{E}}}(P_{FA}^{T_{\mathrm{E}}})=1-\mathcal{F}_{\chi^2_p(\boldsymbol{x}^T\boldsymbol{x})}\left(
\mathcal{F}^{-1}_{\chi^2_p(\boldsymbol{0})}(1-P_{FA}^{T_{\mathrm{E}}})\right) \ .
\label{eq:ED_ROC}
\end{equation}
Indeed, this test does not exploit any prior knowledge on the structure of the object to be detected and can thus be seen as providing a lower  bound for the detection performance.

\subsubsection{Signature of a binary}
\label{sec:TB}
\begin{figure}
\includegraphics[width=8cm]{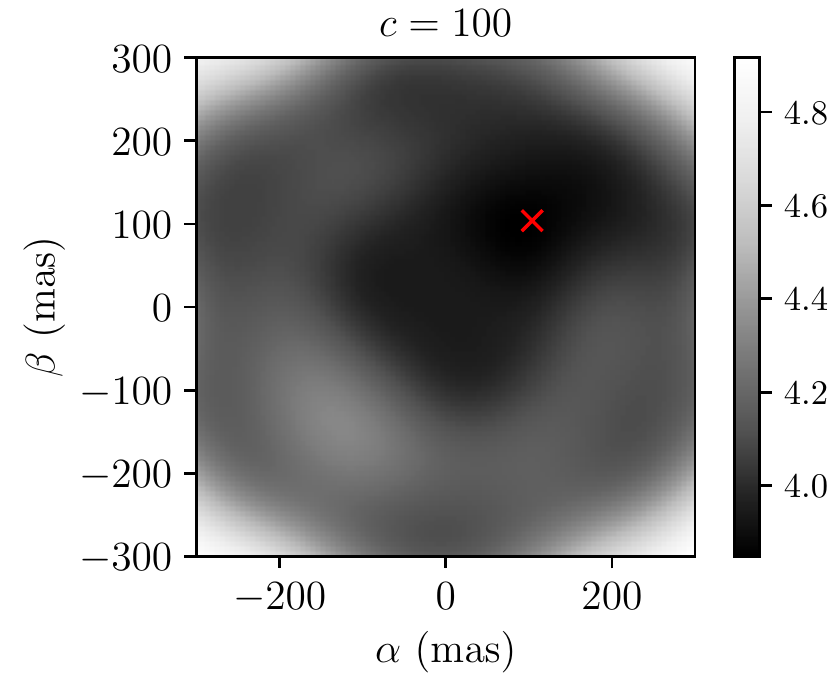}
\caption{Map of the likelihood that is maximised in Eq. \eqref{eq:MLE} for a data vector $\boldsymbol{y}$ accounting for a realistic covariance matrix $\boldsymbol{\Sigma}$ for JWST NIRISS. The companion signature has parameters $\alpha=\beta=104 $ mas (red cross) and $c=100$. }
\label{fig:lmaps}
\end{figure}

Repeated observations of gravitationally interacting multiple systems is the only means by which unambiguous dynamical masses can be determined. Because they make it possible to resolve asymmetries near or even slightly below the diffraction limit, which translates into small orbital distances, NRM closure or kernel-phase \citep{Kraus2008MappingScorpius, Huelamo2011ADisk, Lacour2011SparseVLT} and full-aperture kernel-phase \citep{Pope2013DancingInterferometry, Laugier2019Recovering494AB} are particularly suited to the observation of unequal brightness low mass binary systems.

At any instant, a binary system is characterised by three parameters: the angular separation $\rho$ of the companion relative to the primary, its position angle $\theta$ and a contrast $c$, defined here as the luminosity ratio of the primary over the secondary. Our simulations assume that the position angle is measured in the image relative to the axis pointing up (represented by a blue arrow in the right hand side panel of Fig. \ref{fig:pupil_model}), and increases counterclockwise. Actual observations will also have to take into account the orientation of the telescope to project the apparent position angle onto the celestial sphere to combine observations at multiple epochs.

As an intermediate step, it is also convenient to use a Cartesian coordinate system in which the location of the secondary is given by $(\alpha, \beta)$. If the binary system is made of two individually unresolved point sources, its intensity distribution $\mathcal{O}$ can  be modelled as a pair of Dirac distributions:
\begin{equation}
\mathcal{O}(x,y)=\delta(x,y)+c^{-1}\delta(x-\alpha,y-\beta).
\label{eq:object}
\end{equation}
The complex visibility $\boldsymbol{{\bf v}}$ associated to this object is the 2D Fourier transform of Eq. \eqref{eq:object} \citep{vanCittert1934DieEbene, Zernike1938TheProblems}, that is,

\begin{equation}
\boldsymbol{{\bf v}}(\boldsymbol{u},\boldsymbol{v})=1+c^{-1}\mathrm{exp}\left({\displaystyle{-i\frac{2\pi}{\lambda}(\alpha \boldsymbol{u} +\beta \boldsymbol{v})}}\right).
\label{eq:vis_binary}
\end{equation}

Recall that in the alternative hypothesis defined in Eq. \eqref{eq:H0H1NP}, $\boldsymbol{x}=\boldsymbol{\Sigma^{-\frac{1}{2}}}\boldsymbol{K} \boldsymbol{\phi_0 }$ where $\boldsymbol{\phi_0 }=  \angle \boldsymbol{\bf v }$. This leads to the parametric hypothesis:

\begin{equation}
\mathcal{H}_1: \boldsymbol{y}=\boldsymbol{\Sigma}^{-\frac{1}{2}}\boldsymbol{K}\angle{\left(1+c^{-1}\mathrm{exp}\left({\displaystyle{-i\frac{2\pi}{\lambda}(\alpha\boldsymbol{u}+\beta\boldsymbol{v})}}\right)\right)}+\boldsymbol{\epsilon}.
\label{eq:H0H1binary}
\end{equation}

Under $\mathcal{H}_1$, there are three free parameters: $\alpha$, $\beta$ and $c$, so the MLE is now 
\begin{equation}
\begin{split}
\widehat{\boldsymbol{x}}:&=\underset{\alpha,\beta,c}{\mathrm{argmax}\ } \ell(c,\alpha,\beta;\boldsymbol{y})\\
=&\underset{\alpha, \beta, c}{\mathrm{argmax}\ }{\mathrm{e}^{\displaystyle{-\frac{1}{2}\|\boldsymbol{y}-\boldsymbol{\Sigma}^{-\frac{1}{2}}\boldsymbol{K}\angle{(1+c^{-1}\mathrm{e}^{\displaystyle{-i\frac{2\pi}{\lambda}(\alpha\boldsymbol{u}+\beta\boldsymbol{v})}})}\|^2}}}.\\
\end{split}
\label{eq:MLE}
\end{equation}
Finding the MLE is  equivalent to minimising the argument of the exponential. This minimisation cannot be done analytically but numerical methods can be used to compute $\widehat{\boldsymbol{x}}$, as explained below. Injecting \eqref{eq:MLE} in \eqref{eq:GLR} gives the test
\begin{equation}
\frac{\mathrm{exp}\left({\displaystyle{-\frac{1}{2}(\boldsymbol{y}-\widehat{\boldsymbol{x}})^T(\boldsymbol{y}-\widehat{\boldsymbol{x}})}}\right)}{\mathrm{exp}\left({\displaystyle{-\frac{1}{2}\boldsymbol{y}^T\boldsymbol{y}}}\right)} \underset{\mathcal{H}_0}{\overset{\mathcal{H}_1}{\gtrless}} \eta \ ,
\label{eq:BGLR}
\end{equation}
equivalent to
\begin{equation}
T_{\mathrm{B}}(\boldsymbol{y}):=2\boldsymbol{y}^T\widehat{\boldsymbol{x}}-\widehat{\boldsymbol{x}}^T\widehat{\boldsymbol{x}} \underset{\mathcal{H}_0}{\overset{\mathcal{H}_1}{\gtrless}} \xi \ .
\label{eq:TB}
\end{equation}
Note that this detection problem is similar to the case VII of \citet{Scharf1994MatchedDetectors}, where the detection procedure also relies on the ML estimation of the signal of interest. In that reference, however, the signature $\boldsymbol{x}$ is assumed to live in a linear subspace (independent from the nuisance subspace), which is not the case here.

As mentioned above, the MLE $\widehat{\boldsymbol{x}}$ must be found  numerically. Fig. \ref{fig:lmaps} illustrates, for one realisation of $\boldsymbol{\epsilon}$, an example of the value of the likelihood for a fixed contrast as a function of  position angles $\alpha$ and $\beta$. It is apparent that the likelihood function is multimodal, so  the minimisation strategy must be able to avoid local minima. A brute force search on a finely discretised grid of the parameter space is possible but comes at a large computation cost.
Efficient numerical methods for solving multimodal problems exist, such as {for instance} Monte Carlo Markov Chains method with simulated annealing \citep{Andrieu2003AnLearning} or nested sampling \citep{Skilling2004NestedSampling}.

Because the distribution of $T_{\mathrm{B}}$ involves the unknown distribution of the MLE estimate $\widehat{\boldsymbol{x}}$, it cannot be characterised analytically. However,
as we shall see in the next Section, this distribution can be estimated by Monte Carlo simulations, allowing to establish accurately the relationship between the false alarm probability $P_{FA}^{T_{\mathrm{B}}}$ of this test and the threshold $\xi$ in \eqref{eq:TB}.\\
As an important final remark, we underline that
the false alarm probabilities of the considered tests are independent of the power of the phase perturbations $\boldsymbol{\varphi}$ (at least as long as the linear model in Eq. \eqref{eq:Aphi} holds, that is, for phase perturbation below $\approx 1$ radian). This is clear from  expressions \eqref{pfanp} and \eqref{eq:ED_PfaPdet} for tests $T_{\mathrm{NP}}$ and $T_{\mathrm{E}}$; this is also the case for test $T_{\mathrm{B}}$ because the phase perturbation is cancelled by the operator $\bf{K}$ and does affect the test statistic. This means that the false alarm rate of these tests remains constant in case of fluctuating aberrations, which is a  desirable feature in practice.

\subsubsection{Likelihoods, likelihood ratios and $\chi^2$ intervals}
The test statistic $T_B$ can be interpreted in terms of $\chi^2$-derived intervals as follows. 
Let $\hat{\boldsymbol{x}}$ be some model obtained by
some fit on data $\boldsymbol{y}$. The  $\chi^2$
 score corresponding to this fit is
 \begin{equation}
T_{\chi^2}(\hat{\boldsymbol{x}},\boldsymbol{y}):=\sum_{k=1}^N(\hat{\boldsymbol{x}}_k-\boldsymbol{y}_k)^2=(\hat{\boldsymbol{x}}-\boldsymbol{y})^T(\hat{\boldsymbol{x}}-\boldsymbol{y}).
\label{eq:chi_2}
\end{equation}
Considering the likelihood in Eq. \eqref{eq:def_likelihood},
this shows that if $\boldsymbol{y}$ is Gaussian with mean
$\hat{\boldsymbol{x}}$, the score in Eq. \eqref{eq:chi_2}
 is indeed a $\chi_p ^2$ random variable. Now, the
 test statistics $T_B$ can be rewritten as
 \begin{eqnarray} 
T_B&=& 2\boldsymbol{y}^T\hat{\boldsymbol{x}}-\hat{\boldsymbol{x}}^T\hat{\boldsymbol{x}} = \boldsymbol{y}^T\boldsymbol{y} -\left((\hat{\boldsymbol{x}}-\boldsymbol{y})^T(\hat{\boldsymbol{x}}-\boldsymbol{y})  \right)\\
&=& T_{\chi^2}(\boldsymbol{0},\boldsymbol{y})-T_{\chi^2}(\hat{\boldsymbol{x}},\boldsymbol{y}),
\label{eq:chi2_TB}
 \end{eqnarray}
which shows that $T_B$ can be interpreted as
the reduction in the sum of squared residuals when
comparing the null hypothesis to the considered model.\\
For the sake of accurately controlling the false alarm rate, note, however, that $T_{\chi^2}$ in Eq. \eqref{eq:chi_2} may not be distributed as a ${\chi^2_p}$
variable because $\hat{\bf{x}}$ is a random variable. Actually, the true distribution of  $T_{\chi^2}$ 
may not be known analytically, and a Monte Carlo procedure (such as
that mentioned in Sec. \ref{sec:TB}  for the estimation of the correspondence between the $P_{FA}$ vs threshold for $T_B$) is required.

\section{Results}
The  tests with the performance analyses presented in Section \ref{sec:ker_stats} are very general: considering a different aperture and instrumental noise simply amounts to replacing $\bf{A}$,  $\bf{K}$ and
$\bf{\Sigma}$ in the equations.
We focus now on their specific application to JWST NIRISS full pupil images.
\label{sec:results}

\subsection{Dataset and considered targets}
\label{sec:JWST_analysis}

We will apply the three detection tests previously introduced to a series of simulated JWST/NIRISS datasets, replicating the observing scenario of archetypal ultracool Y-type brown dwarfs. While their multiplicity rate is currently unknown, over 25 such objects have been discovered less than 20 pc away, mostly by the WISE mission \citep{Kirkpatrick2011TheWISE}. At 20 pc, the theoretical angular resolution of JWST for $\lambda = 4.8\; \mu$m translates into an orbital distance of 3 AU: interferometric observations will make it possible to probe within the first few AUs of most known Y dwarfs.

JWST NIRISS images of Y dwarfs are simulated to evaluate the performance of the detection tests, using the \texttt{ami\_sim}\footnote{\texttt{ami\_sim} is available at https://github.com/agreenbaum/\\ami\_sim} package \citep{Greenbaum2016Ami_sim}, corresponding to a 40-minute integration on target and a 40-minute integration on a perfect calibrator. Frames are simulated in full pupil mode, using the F480M filter, for two different `W2' magnitudes: 15.4 and 14.1. The W2 magnitude is the apparent magnitude in the band selected by the W2 ($\lambda = 4.6\;\mu m$) WISE filter \citep{Wright2010ThePerformance}. For these objects, companions are placed at a single position angle $\theta=315^{\circ}$ (materialised by the orange arrow in the PSF shown in Fig. \ref{fig:pupil_model}). The simulated companions lie at separations of $\rho=73$ mas ($\approx0.5\lambda/D$ @ $\lambda=4.8\mu$m) or $\rho=147$ mas ($\approx\lambda/D$ @ $\lambda=4.8\mu$m), and have contrasts $c=10$, $c=20$, $c=50$ or $c=100$, leading to a total of 8 possible signatures.

For any given target, a calibration frame is simulated and we assume no calibration error (stable wavefront, calibrator with the same spectrum and brightness as the Y dwarf). To comply with a real situation, ker-phases are not extracted directly from the simulated image: the frames are recentered, cropped to a size of $64\times64$ pixels and apodized by a Super-Gaussian mask (see Eq. 2 of \citet{Laugier2019Recovering494AB}) of radius 30 pixels to weigh down the edges of the image.

\subsection{Modelling the errors}
\begin{table}
\small
\centering
\begin{tabular}{|c|c|}
\hline 
Read noise ($e^-$)			&14.849			\\
Flat field error 			&$0.01\%$		\\
Dark current ($e^-/s$)		&$0.04$			\\
Total integration time (s)	&$2400$			\\
Number of frames 			&$15$			\\
Gain (e$^-$/ADU)            &$1.00$         \\
Jitter value (mas)          &$7.0$          \\
Integration time    &$40$ minutes   \\
\hline 
Number of photons (W2 mag = 15.4)		&$3.723\times10^{6}$	\\
Number of photons (W2 mag = 14.1)		&$1.1181\times10^{7}$	\\
\hline
\end{tabular}
\centering
\vspace*{3mm}
\caption{Detector and targets characteristics used to compute the covariance of the ker-phases extracted from our JWST NIRISS simulated dataset.}
\label{tab:sim_images}
\end{table}
\label{sec:errors}

Two types of errors affect kernel-phases and the outcome of the statistical tests described in Sec. 2. First are statistical errors induced by random noises whose overall impact can be captured in the acquisition or the synthesis of a global covariance matrix. Second are systematic errors resulting from the imperfect modelling by the kernel framework of the broadband, long-exposure, and diffractive nature of images. The subtraction of kernel-phases acquired on a point source theoretically accounts for this systematic error however, in practice, wavefront drifts between observations will result in unaccounted residual residual errors referred to as systematic errors \citep{Ireland2013PhaseMasking}.

To estimate the potential impact of wavefront drift induced systematic errors, we rely on \citet{Perrin2018UpdatedStrategies} who predict that over a timescale of two hours, JWST drifts will result at most in a 16 nm RMS wavefront across the entire pupil\footnote{\citet{Perrin2018UpdatedStrategies} predict that large variations in slew angle will result in the most important variations, as the primary mirror regains thermal equilibrium over the course of days.}.
We used the 10 OPD maps distributed with the \texttt{webbpsf} package, scaled down to correspond to the predicted RMS to produce images resulting in 10 distinct ker-phase realisations. The dispersion of ker-phases across these realisations was used to estimate the magnitude of the calibration residual. In the bright target scenario (W2 mag = 14.1) introduced in Sec. 3.1, this calibration residual accounts for about 14 \% of the total noise variance. As will be shown later, this systematic error has a small impact when observing faint targets.

\subsection{Covariance estimation}
\label{covest}
\begin{figure}
\centering
\includegraphics[width=9cm]{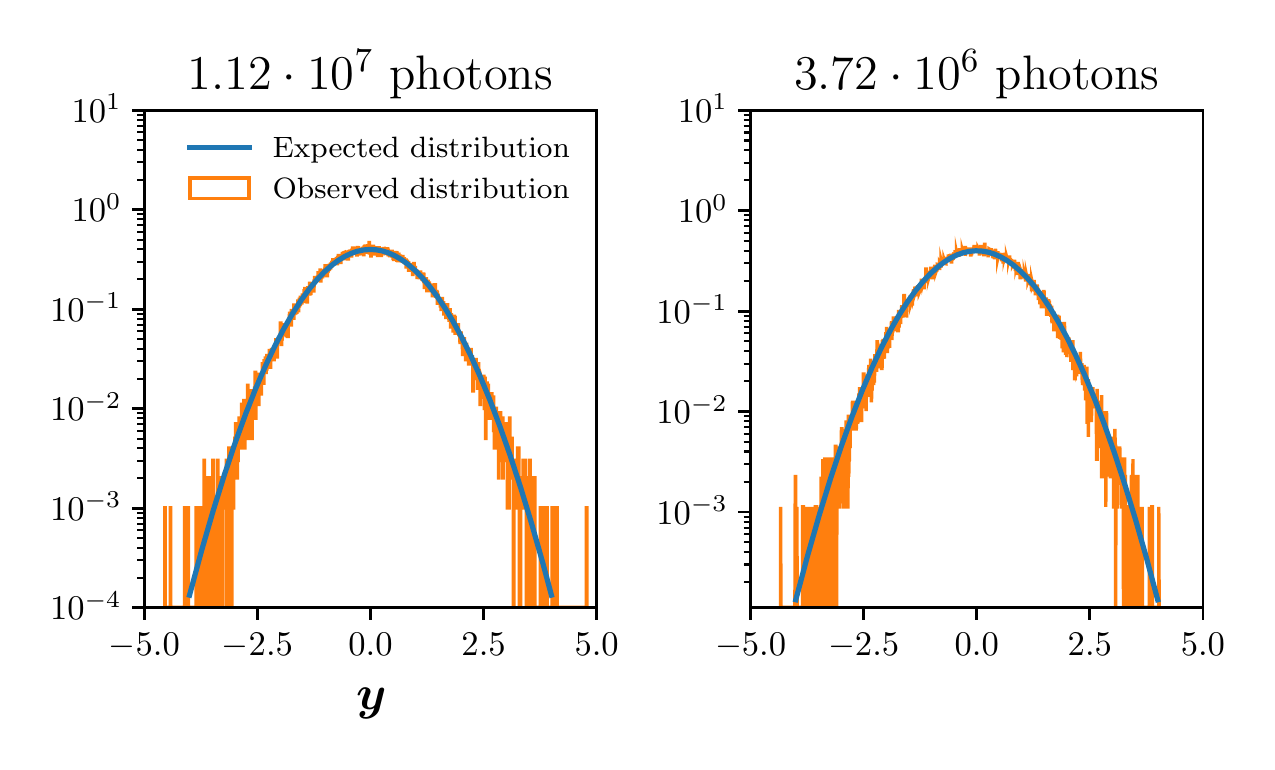}
\centering
\caption{Orange: Histogram of the values of the whitened ker-phases for the calibration images. Blue: standard, normal distribution. Left-panel: higher flux regime. Right panel :lower flux regime. The distribution of whitened ker-phases obtained in practice is accurately described by the theoretical normal distribution considered in Eq. \eqref{eq:H0H1xy}.}
\label{fig:normality}
\end{figure}
\begin{figure}
\centering
\includegraphics[width=9cm]{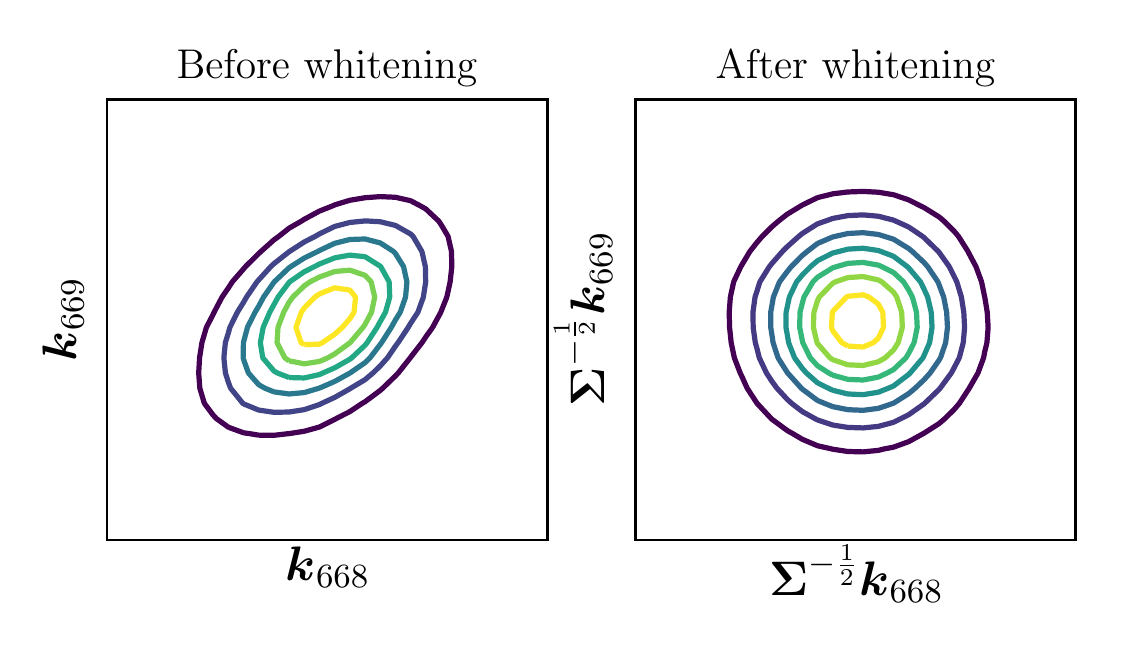}
\centering
\caption{Isocontours of the  joint probability density function (pdf)} of the 668$^{\textrm{th}}$ and 669$^{\textrm{th}}$ elements of ker-phases vector $\boldsymbol{k}$,  $\boldsymbol{k}_{668}$ and $\boldsymbol{k}_{669}$, before  (left) and after (right) whitening. The PDFs are estimated using $10^5$ noise realisations.
\label{fig:whitening}
\end{figure}
\label{sec:est_cov}

Whereas simulated images used in the analysis include all the previously listed noises, experience has shown us that, barring calibration residuals, the covariance matrix can accurately be estimated using the three dominant noises: photon, readout  and dark current. Fig. \ref{fig:normality} indeed shows that after whitening by this simpler covariance, the distribution of kernel-phases is indistinguishable from a normal distribution of standard deviation 1.

The effect of the whitening is further illustrated in Fig. \ref{fig:whitening}, which shows how previously noise-correlated kernel-phases (left panel) are indeed made statistically independent (right panel). The thus whitened observables can indeed be reliably used as input for the different statistical tests introduced in Sec. \ref{sec:stat_models}.

In practice, the covariance $\boldsymbol{\Sigma}$ is estimated using Monte Carlo simulations. An accurate estimation requires a number of {simulated} frames much greater than the total number of kernels; we used $10^5$ frames for 887 kernels in our case.

Calibrated ker-phases are obtained by subtracting the ker-phases of a calibrator from the ker-phases of the target in order to remove kernel model imperfections. Since the same flux is assumed for both observations, they share the same covariance. The covariance of the calibrated ker-phase vector is therefore twice the covariance $\boldsymbol{\Sigma_{est}}$ estimated from the MC simulations.

To account for unknown calibration errors reported in NRM-inteferometry as well as in full aperture kernel-phase that result in a kernel-phase bias, one commonly used solution has been to artificially inflate the experimental variance by adding an additional term whose overall magnitude is adjusted during the model fit (eg. \citet{Martinache2009ApertureSubaru}). The OPD maps introduced in Sec. 3.2 make it possible to estimate the magnitude of this bias \textit{a priori}. Proper treatment of the calibration would require the subtraction of an estimate of the calibration term, using either the POISE algorithm of Ireland (2013) or the KL decomposition approach described by \citet{Kammerer2019KernelLimit} that relies on the observation of multiple calibration sources. Here we will estimate the impact of an unaccounted calibration error on the contrast detection limits by adding the residual determined after analysis of the simulation that included the OPD maps to the diagonal of the covariance. To pursue the possibly covariated effects would require the computation of a distinct covariance matrix from a large number of distinct realisations of telescope drifts. For the faint brown dwarf case that motivates this study, the impact of the calibration error are small, so we chose not to pursue the non-diagonal terms.\\

\subsection{Parameter estimation}
\label{sec:params}
\begin{figure}
\centering
\includegraphics[width=9cm]{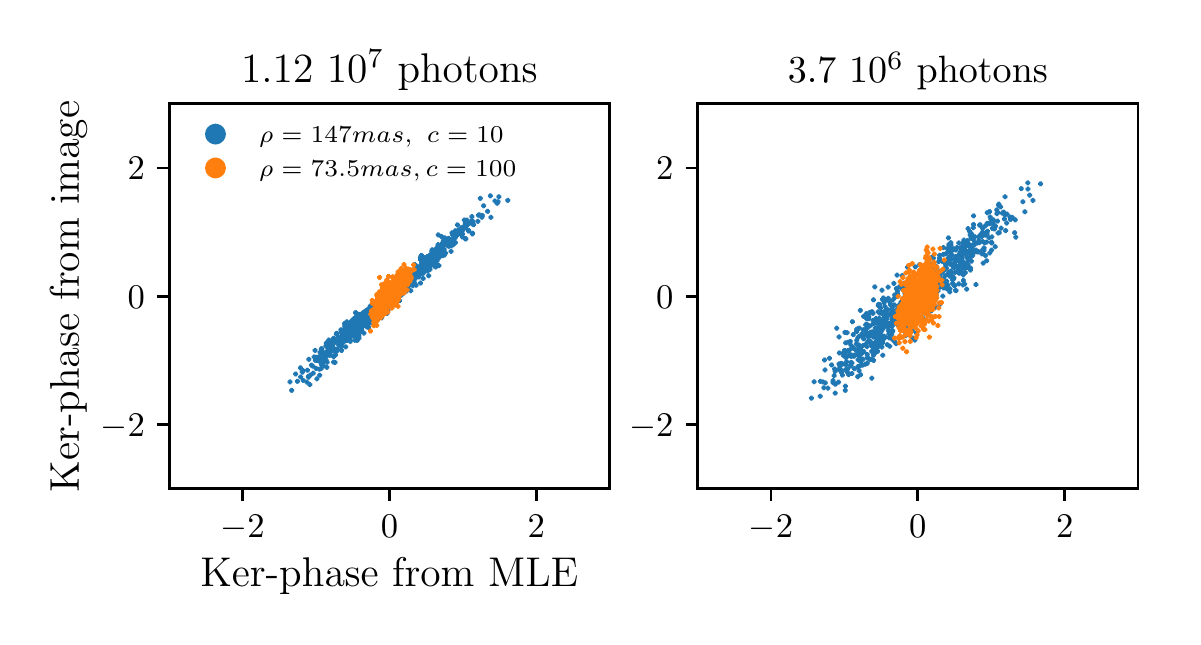}
\centering
\caption{Ker-phases of the recovered signature ($x$ axis) against the true ker-phases of the injected binary ($y$ axis). Left panel: high flux regime, right panel: low flux regime. The worst SNR situation ($\rho = 73.5 mas,\ c=100$) is in orange, and the best SNR ($\rho = 147 mas,\ c=10$) in blue.} 
\label{fig:recovered_vs_injected}
\end{figure}
\begin{figure}
\centering
\includegraphics[width=9cm]{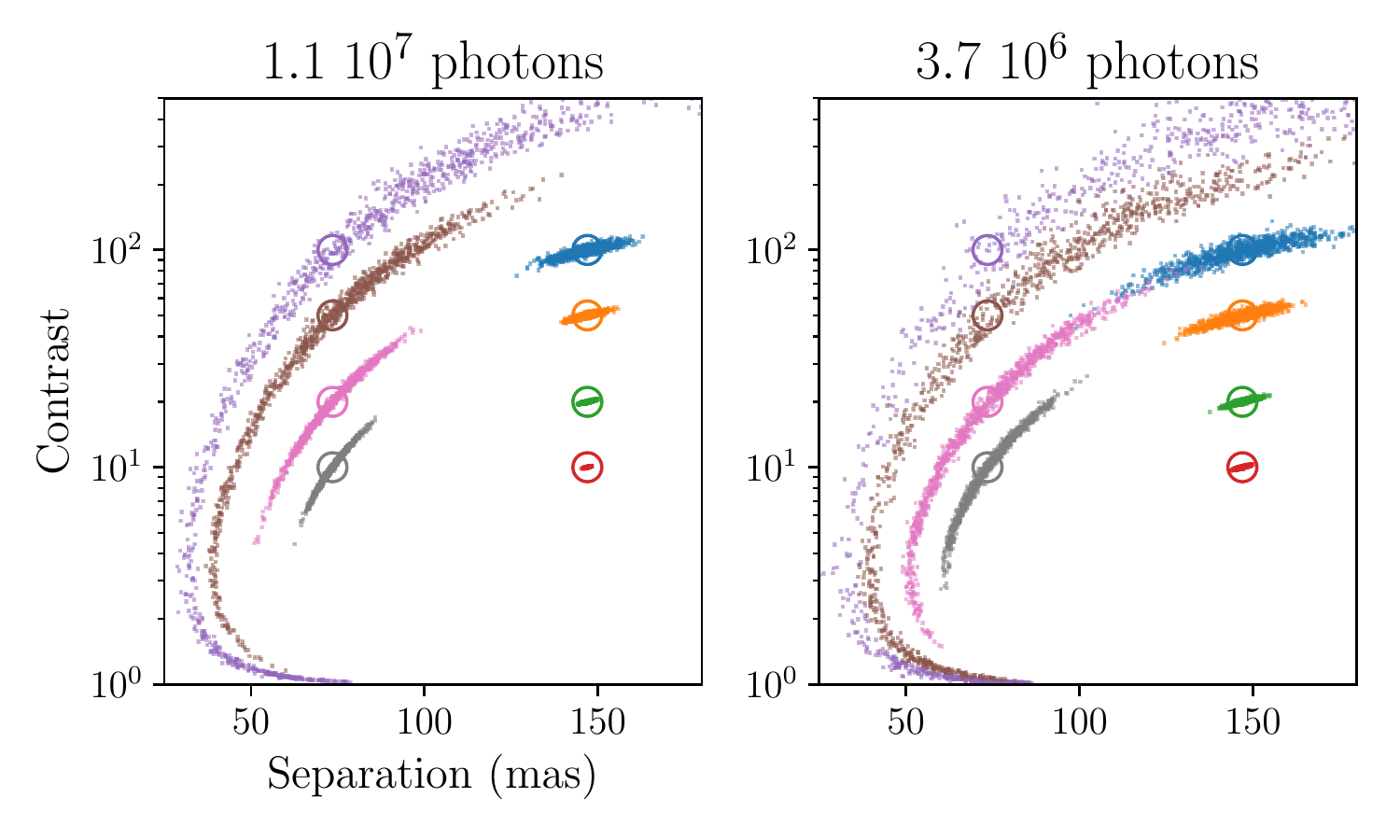}\\
\hspace*{0.3cm}\includegraphics[width=4.25cm]{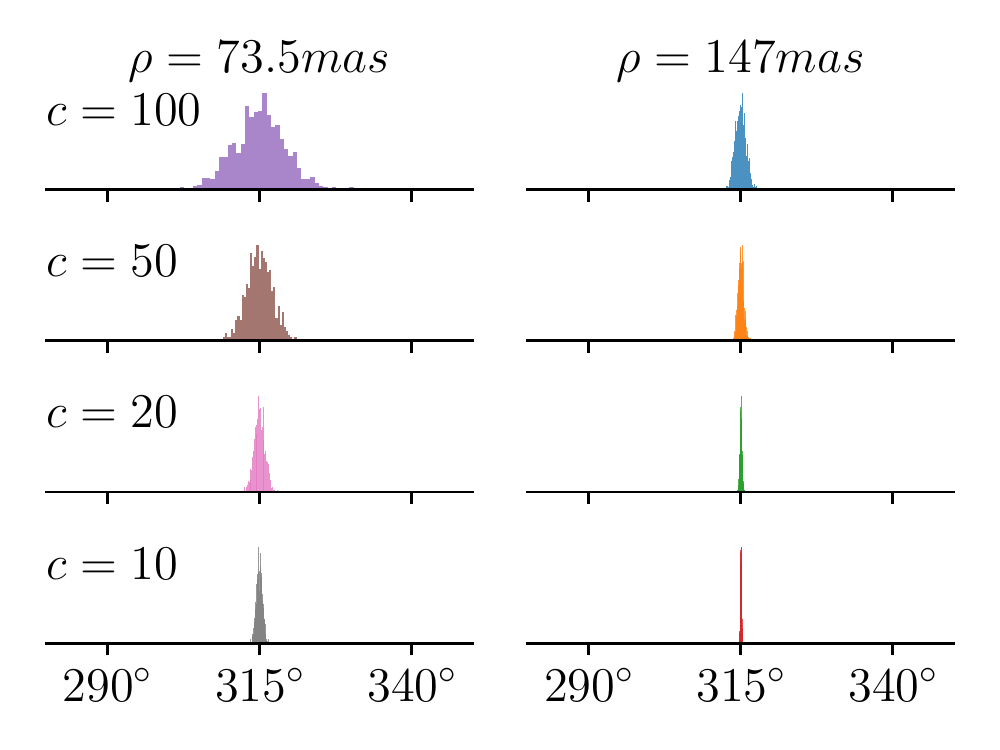}~
\includegraphics[width=4.25cm]{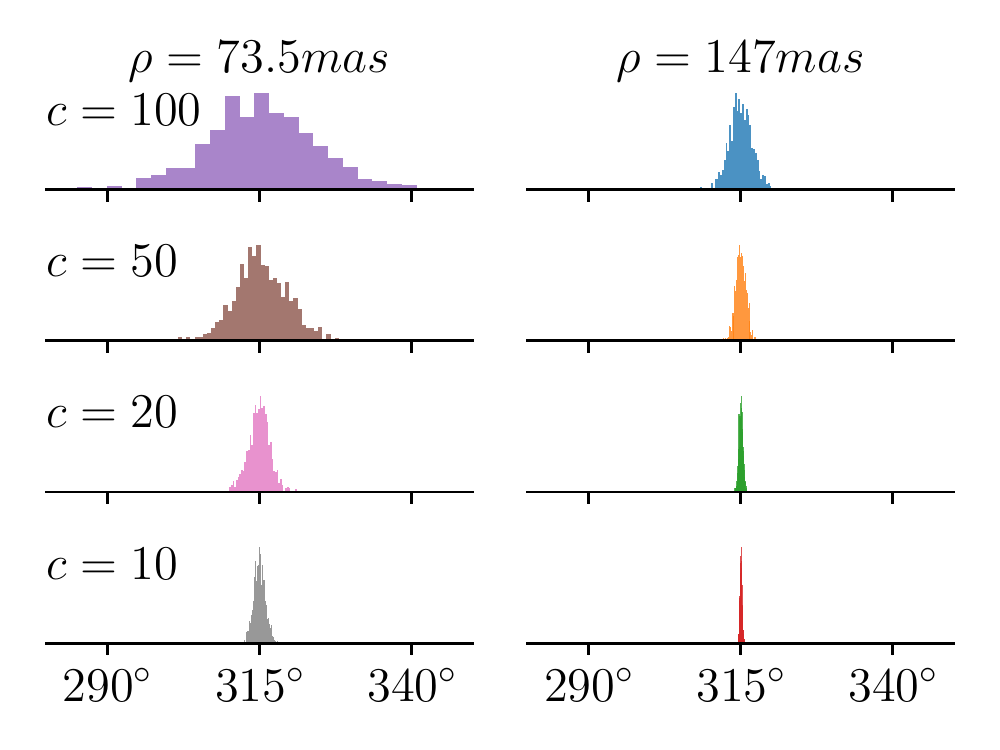}
\centering
\caption{Error on the recovered parameters. Top panels: the circles represent the separation and contrasts of the injected signature, and each column represent one flux regime. The bottom panels represent the error on the estimated angle $\theta$. The parameters $\rho$ and $c$ of each injected signature are represented as coloured circles on the top panels, while $\theta$ is fixed at $315^{\circ}$ for all signatures. The same colour code is used for every panel, with each colour corresponding to an injected signature. Each dot on the top panel represents the parameters estimated for a single realisation of the noise.}
\label{fig:errors}
\end{figure}
Detecting a companion using the operational binary test $T_B$ requires the determination of the MLE $\widehat{\boldsymbol{x}}$. This requires estimating the parameters $\rho$, $\theta$ and $c$ from the whitened ker-phases $\boldsymbol{y}$ in Eq. \eqref{eq:H0H1binary}. The distribution of the parameters can be estimated by generating, for each considered signature, a large number of noisy ker-phases and estimating the parameters. In practice, a global optimisation algorithm can be used. For the purpose of making a large number of simulations, we assume that the algorithm has localised the region in which the global minimum is situated (the darkest region in  Fig. \ref{fig:lmaps}). In this setting, the minimum can be found by a gradient descent algorithm. 

In the following, we use the algorithm described by \citet{Branch1999AProblems}, as implemented in \texttt{scipy.optimize.leastsquares}, which uses the local gradient and  optimises for the direction descent and step size. The initialisation of the algorithm corresponds to the parameters of the injected companion. This method is suited for the determination of contrast limits thanks to its speed. We checked that we obtained very similar results with a (computationally more expensive) systematic grid search that would typically be used in practice. \footnote{The gradient descent procedure is indeed only applicable in the context of only applicable in the context of the determination of detection limits by a Monte Carlo method.}

Fig. \ref{fig:recovered_vs_injected} shows the recovered ker-phases as a function of the ker-phases of simulated images for different separation, contrast and flux regimes. The fit remains pretty consistent for each case, with a scatter getting predictably more important as the signal-to-noise ratio decreases (the signal-to-noise ratio is affected by the contrast, the separation and the total flux in the image).

All of the signatures presented in Fig. \ref{fig:errors} are detectable by the  $T_B$ with $P_{FA}<10^{-3}$. The shape of the two dimensional distribution of the estimated separation $\rho$ and contrast $c$ reproduces what was for instance reported by \citet{Pravdo2006Masses231.1BC} in the context of NRM observations: at angular separations smaller than $\lambda/D$, estimates for the contrast and the angular separation are strongly correlated.

Fig. \ref{fig:errors} also shows that two regimes can be distinguished. For a companion at $\rho \approx \lambda /D$ (for JWST $\lambda/D=152$mas @ $\lambda4.80\mu$m), all parameters are well constrained, while for a companion at $\rho < \lambda/D$, the contrast and the angular separation cannot be well constrained simultaneously. In practice, this means that the estimation of the position of a companion using kernel-phases when the expected angular separation is smaller than $\lambda/D$ can be further constrained by an independent measurement of the luminosity of the companion at a different epoch, when $\rho > \lambda / D$. This property can be particularly useful in the case of objects with high eccentricities or inclinations.\\
A study of the consequences of the parameters' uncertainties and correlations on the orbit that can be fitted using the Kernel method on NIRISS images is out of the scope of this paper; this should be the object of future work, along with  recommendations of optimum observing strategies in regards of the uncertainties on measured orbital parameters.
\subsection{Detection and contrast performance}
\label{sec:perfs}
\begin{figure}
\centering
\includegraphics[width=7cm]{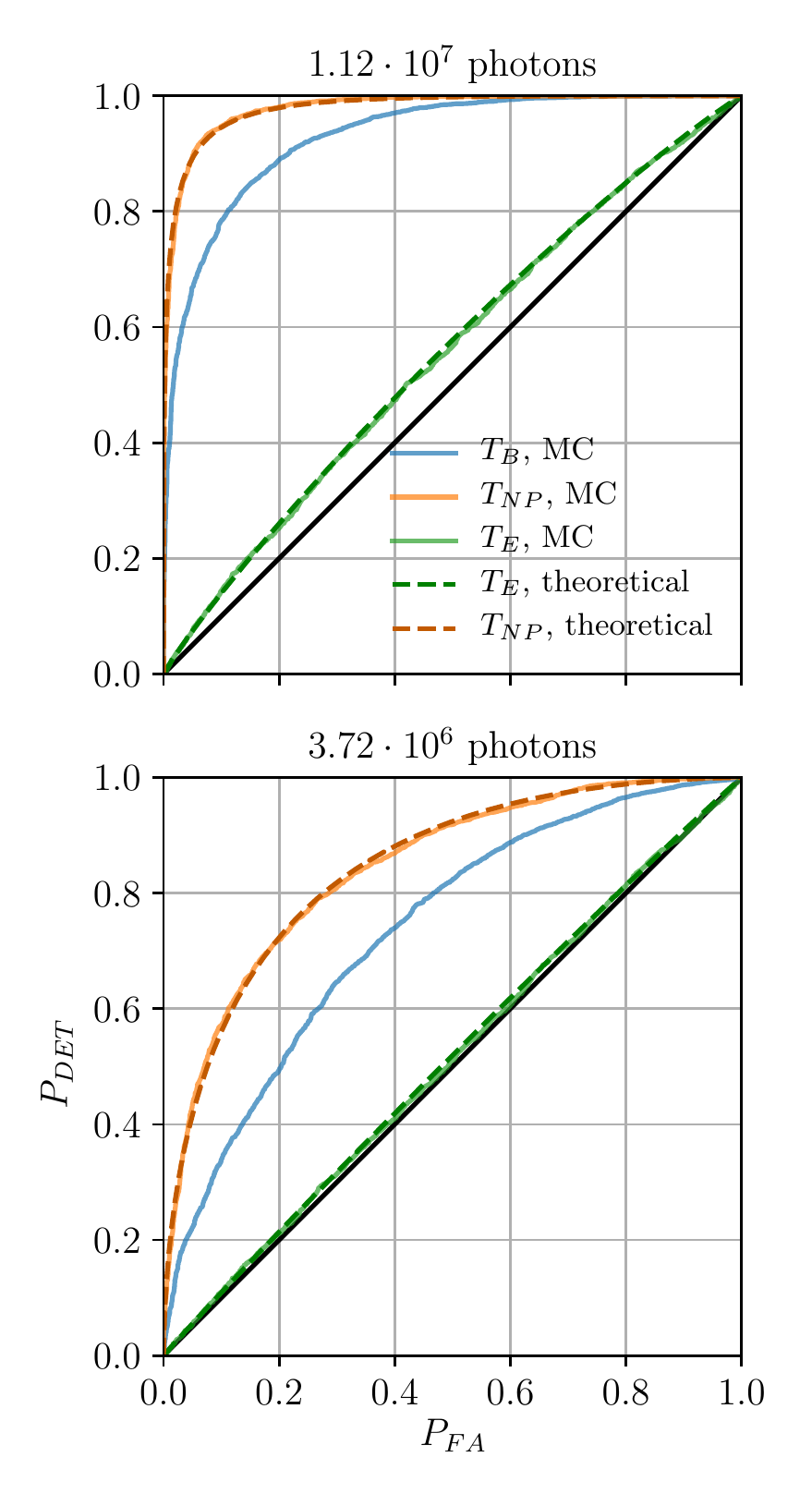}
\centering
\caption{ROC curves of $T_E$ (green), $T_{NP}$ (blue) and $T_B$ (orange). Theoretical ROC curves for $T_{NP}$ and $T_E$ plotted using Eq. \eqref{eq:NP_PfaPdet} and Eq. \eqref{eq:ED_PfaPdet}, for a companion at $\rho=200$ mas, $c=1200$ and $\theta=45^{\circ}$ off the vertical. Dashed lines correspond to theoretical ROCs, while solid lines represent ROCs obtained by Monte-Carlo simulations.} The closer a curve is from the black line on the diagonal, the less powerful the corresponding test. The higher flux regime is represented in the top panel, and the lower flux regime in the bottom panel. The performance of $T_{NP}$ and $T_E$ are accurately described by the theoretical expressions in Eq. \eqref{eq:NP_PfaPdet} and Eq. \eqref{eq:ED_PfaPdet}. The test $T_{NP}$ presents the highest performance. $T_B$ is the next best-performing test and $T_E$ has the lowest performance of the three. We see a clear improvement of the power of all tests as the flux (and thus the SNR) increases.
\label{fig:ROC}
\end{figure}
\begin{figure}
\makebox[\linewidth][c]{\includegraphics[width=9cm]{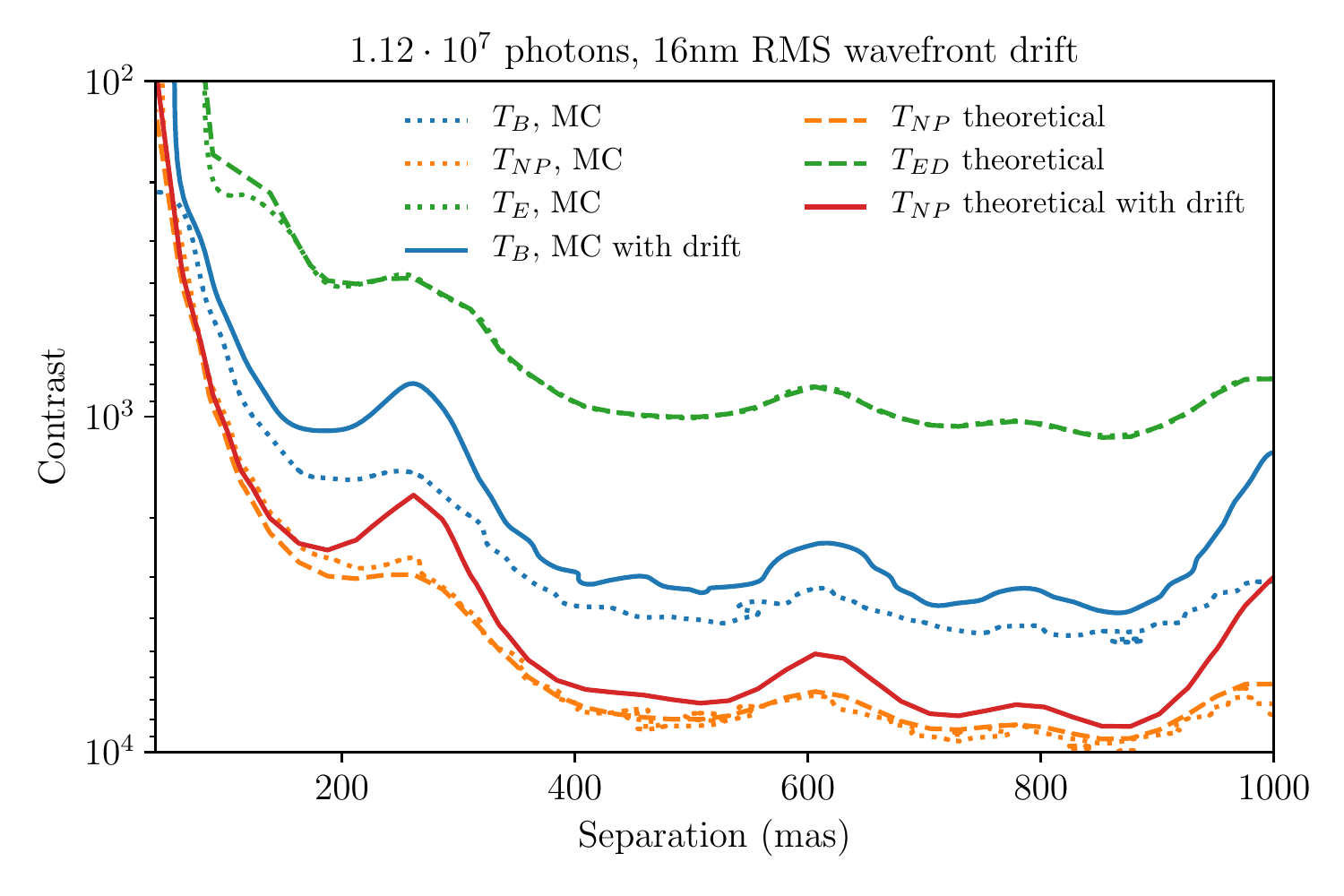}}
\caption{ Detection limits at a fixed position angle $\theta=315^{\circ}$: contours below which the  $P_{DET}$ falls below $68\%$ for a fixed $P_{FA}$ of $1\%$, represented as a function of the separation and contrast of the companions, for $T_E$ (green), $T_B$ (blue) and $T_{NP}$  (orange).
The dashed lines represent theoretical detection limits for $T_E$ and $T_{NP}$ (Eq. \eqref{eq:NP_PfaPdet} and Eq. \eqref{eq:ED_PfaPdet}) and the dotted lines present the limits actually achieved in the MC simulations. $T_{NP}$ (orange) provides ideal detection limits for a Kernel treatment of a JWST-NIRISS image and the practical test $T_B$  (dotted blue)  has contrast detection limits within a factor of 2.5 of the theoretical maximum.} The solid lines represent the detection limits for $T_B$ (blue) and $T_{NP}$ (red) with a calibration residual corresponding to a 16~nm RMS wavefront drift.
\label{fig:det_lims}
\end{figure}
\begin{figure}
\centering
\includegraphics[width=9cm]{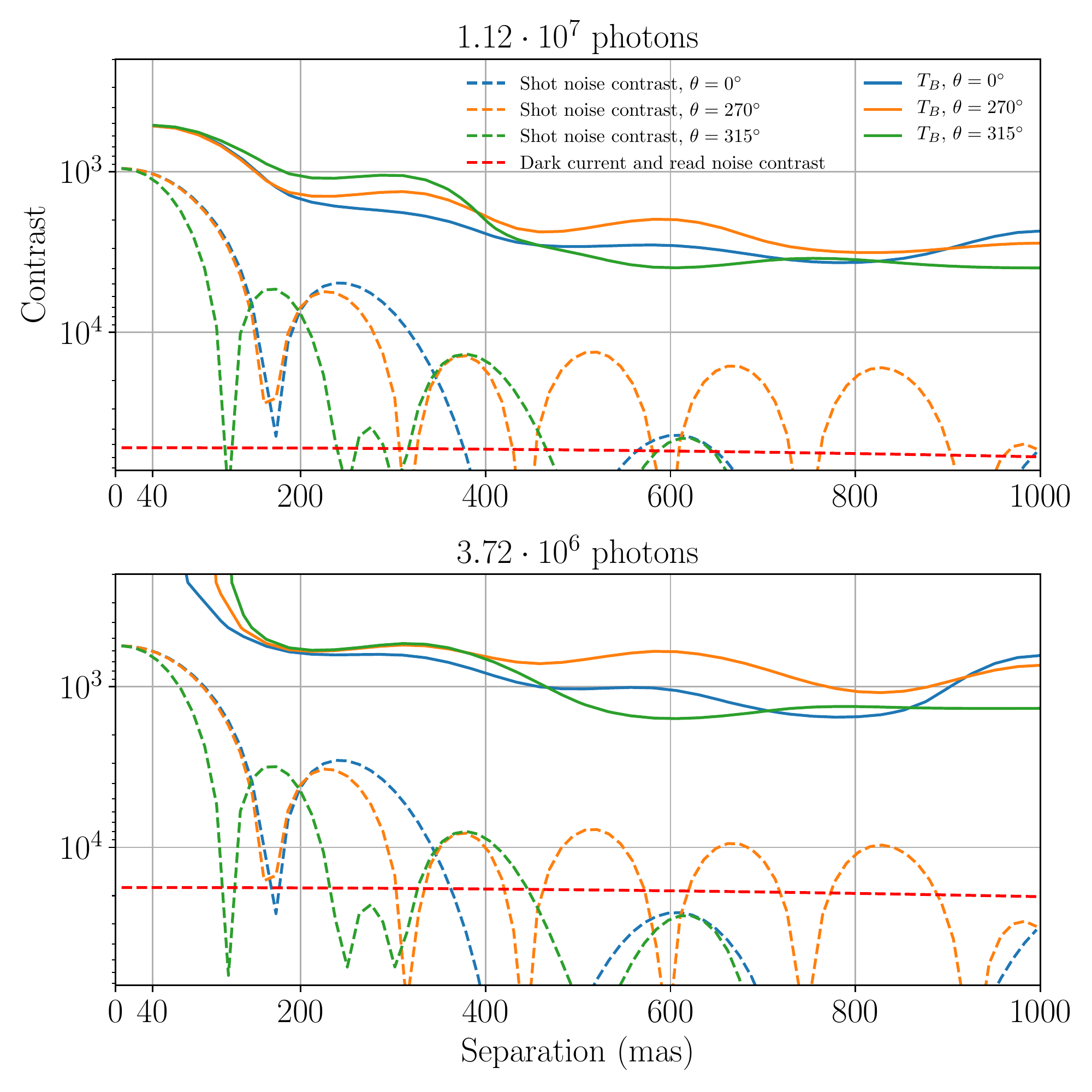}
\caption{Detection limits for test $T_B$ (Eq. \eqref{eq:TB}), in the higher flux regime (top panel) and the lower flux regime (bottom panel). The solid lines correspond to contours of $P_{DET}=68\%$ at a fixed $P_{FA}=1\%$. Detection limits are represented at three different position angles for the companion: 0, 45 and 90 degrees off the vertical, as orientated in the PSF shown in Fig. \ref{fig:pupil_model}. The relative signal-to-noise ratios (see text) are indicated by dashed lines. The shot (photon) noise is the main limiting noise in most cases.}
\centering
\label{fig:det_map}
\end{figure}
First, we validate the theoretical relations predicting the performance of the NP test $T_{NP}$ (Eq. \eqref{eq:NP_PfaPdet}) and of the energy detector $T_E$ (Eq. \eqref{eq:ED_PfaPdet}), and determine the actual performance of $T_B$ (Eq. \eqref{eq:TB}). For that purpose, we perform Monte Carlo simulations consisting in 2000 realisations\footnote{The number of realisations is dictated by the target $P_{FA}$ and $P_{DET}$. For the considered $P_{FA}=1\%$ and $P_{DET}=68\%$, $2000$ realisations correctly sample the distributions of the test statistic of $T_B$ under $\mathcal{H}_0$ and $\mathcal{H}_1$.} of $\boldsymbol{y}$ under $\mathcal{H}_0$ and under $\mathcal{H}_1$ for a given signature $\boldsymbol{x}$ (cf Eq. \eqref{eq:vis_binary}). 

All of the detection limits are shown for $P_{FA}=1\%$ and $P_{DET}=68\%$. In terms more frequently encountered in astronomy publications, this is equivalent to having a $68\%$ chance of making a $\approx 2.3 \sigma$ detection.

On each realisation, we perform each of the three tests by using the kernels operator $\boldsymbol{K}$ and the covariance matrix $\boldsymbol{\Sigma}$ estimated as in Sec. \ref{covest}.

We present in Fig. \ref{fig:ROC} the results in the form of ROC curves, which provide a graphical representation of the power of each test. It can be seen that the dashed curves representing the theoretical ROCs match accurately the solid lines corresponding to the  performance achieved in practice. As expected, $T_{NP}$  appears as the most powerful of the three tests (this test corresponds to the upper performance bound) and $T_E$ as the least powerful of the three (this test uses no prior information on the target signature and can be seen as a lower bound). The performance of $T_B$ logically lies in between, but much closer to the upper than to the lower bound.

The detection limits for the three  tests $T_{NP}$, $T_E$ and $T_B$ are represented in Fig. \ref{fig:det_lims} across a range of contrasts and separations, for a fixed position angle $\theta=315^{\circ}$. The dashed lines correspond to no wavefront error while the solid lines correspond to $16$ nm RMS of wavefront error. We can see that the theoretical performance, validated for a single companion signature in Fig. \ref{fig:ROC} do hold true over a large range of contrasts and separations, and that the detection limit of $T_B$ remains close  to the bound provided by $T_{NP}$. The dashed and dotted lines correspond to a perfectly stable JWST leading to a perfect calibration of the systematic errors.

The detection limits further depend on $\theta$, because the PSF of JWST NIRISS is not centrosymmetric (as visible in Fig. \ref{fig:pupil_model}). Fluctuations of these limits are shown in Fig. \ref{fig:det_map} for three position angles. The Figure also indicates the signal-to-noise level at the corresponding positions in the image (computed here as the maximal pixel value of a noiseless image with only the companion, divided by the standard deviation of the considered noise), showing that the detection limits follow the overall noise level in the image. Performance wise, the detectable contrast ratios are of the order of $10^{3}$ at $200$ mas, with some variations between the two flux levels considered. 
\subsection{Mass limits for WISE 1405+5534}
\begin{figure}
\includegraphics[width=9cm]{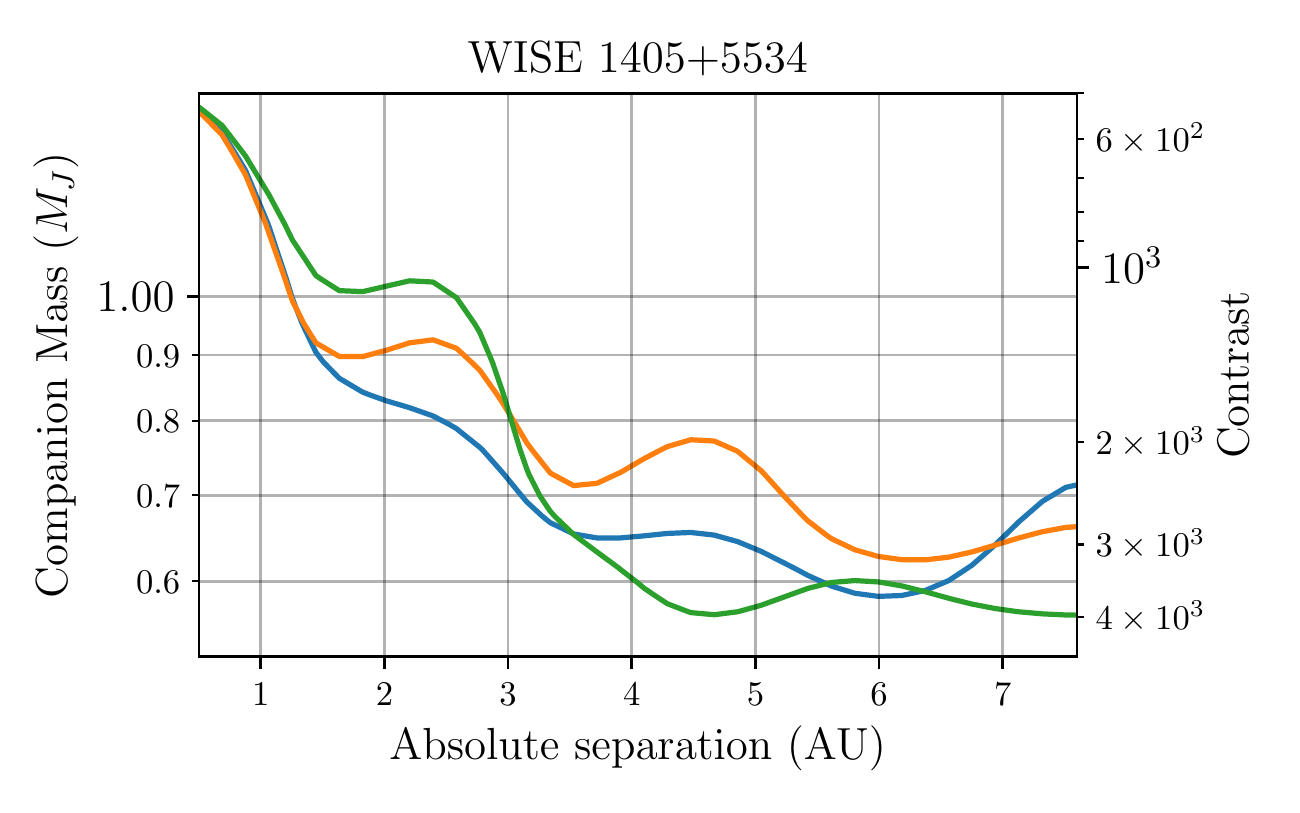}
\caption{Detection limits  of a possible companion to WISE 1405+5534 at $P_{FA}=1\%$ and $P_{DET}=68\%$, as a function of his contrast (right ordinate axis) or mass (left ordinate axis) and absolute separation in AU. {A one Jupiter-mass object is detectable down to $1.5$ AU from the primary}.}
\label{fig:detection_mass}
\end{figure}

WISE 1405+5534 is a Y-type brown dwarf with a W2 magnitude of 14.1 that was used as a reference target to produce the contrast detection limits featured in Fig. \ref{fig:det_lims}. The raw observational detection limit curve of contrast as a function of angular separation can be converted into an astrophysical detection limit curve of companion mass as a function of orbital separation.


Whereas the $129 \pm 19$ mas parallax measured by \citet{Dupuy2013DistancesObjects} directly allows for the conversion of the angular separation into a projected orbital distance, the contrast to mass conversion requires a model. We use the mass-luminosity relations given by the AMES-Cond model of \citet{Baraffe2003Evolutionary209458} for an age of 1 Gy and a mass estimate of 30 M$_J$ for the primary given by \citet{Cushing2011THE}.


The detection limits obtained for WISE 1405+5534 are shown in \ref{fig:detection_mass}. At $P_{FA}=1\%$, and $P_{DET}=68\%$, a $1M_J$ can be detected at separations greater that 1.5 AU. An orbit with this semi major axis would have a period of 40 years, thus a quarter of an orbit could be captured with repeated observations over the expected service life of JWST.

\subsection{Bright limits}
\label{sec:bright}
\begin{figure}
\includegraphics[width=9cm]{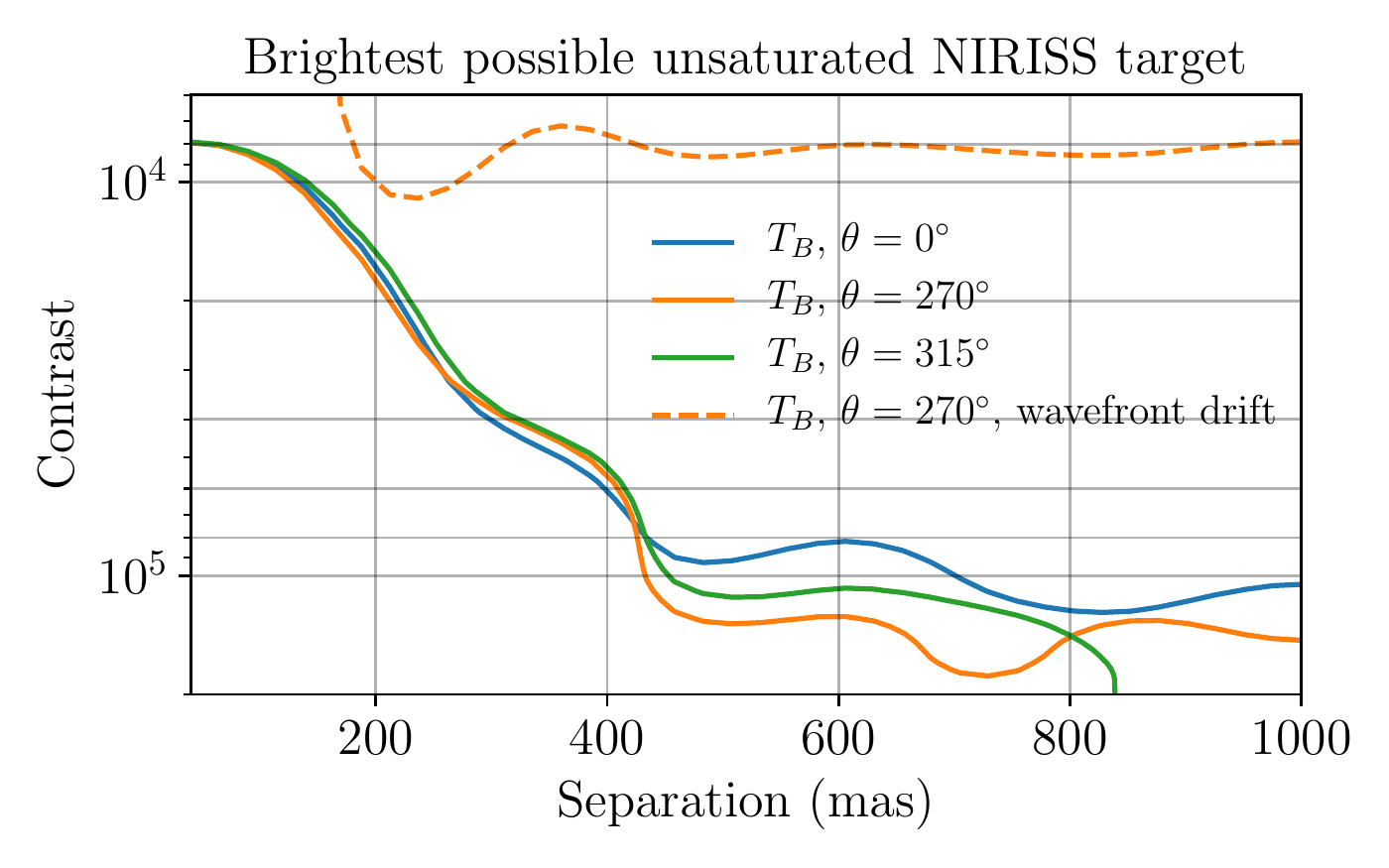}
\caption{Detection limits for the brightest target observable without saturation with JWST NIRISS. Solid lines: detection limits for $T_B$ at $P_{FA}=1\%$ and $P_{DET}=68\%$ applied to the image with the greatest possible dynamic range, with 20 minutes total integration time. For the brightest images, the kernel method with the test $T_B$ ideally allows to detect contrasts up to $10^5$ beyond 500 mas. The dashed orange line represents detection limits in the presence of a 16 nm wavefront drift.
}
\label{fig:detection_deep}
\end{figure}

For the faint Y-dwarf targets considered thus far, it may have occurred to the reader that the contrast detection limits are dominated by the effect of the dark current and the readout noise and not by the photon noise of the central object. We wish here to complete the description of the properties of our approach with a bright target scenario that will feature a different behaviour, thus exhibiting the contribution of the photon noise.

The saturation limit for full pupil JWST NIRISS using the F480M filter, and a 64$\times$64 pixels subarray size is 7.6 mag. We consider a shorter observation sequence, with a total of 20 minutes spent on the target of interest and 20 minutes on a calibrator of similar brightness. 
The detection limits for this observation, using the operational test $T_B$ are shown in Fig. \ref{fig:detection_deep}, at $P_{FA}=1\%$, and $P_{DET}=68\%$.

Unlike the contrast detection limits obtained on the faint targets, the curves now clearly reveal two different regimes. Up to an angular separation of $\approx$500 mas, where the photon noise is expected to dominate, the contrast detection decreases as a function of the angular separation. Beyond this point, it reaches a plateau, as the detection is once again dominated by the homogeneous properties of the dark current and the readout noise.

In this bright scenario, calibration errors induced by a drift comparable to what was described in Sec. \ref{sec:perfs} will have a stronger impact on the weak signal of a high-contrast companion. \citet{Sallum2019ComparingCharacterization} feature contrast detection limits for NIRCam in a similar scenario that takes calibration errors into account. Under the hypothesis introduced in Sec. \ref{sec:errors}, the calibration error accounts here for 85\% of the total noise variance of the kernels and therefore result in a degraded performance by a factor $\approx$10, as shown by the dashed curve in Fig. \ref{fig:detection_deep}.

\section{Conclusion}
This paper provides a theoretical and numerical  analysis of the performance of various detection tests based on the Kernel method. The approach provides an upper bound for the achievable detection limits, and an operational detection test whose performance are close to the upper bound. Furthermore, the false alarm rate of these tests is not affected by fluctuating aberrations and can be tuned a priori.

The kernel-based detection approach presented in this paper is not specific to either NIRISS, the 480M band, the full pupil imaging mode, nor to JWST itself. The method only requires weak wavefront perturbations and appropriate sampling (i.e., a small enough platescale as compared to $\lambda/D$). In particular, the statistical treatment proposed in this study can also be used for NRM data.

For JWST-NIRISS in the F480M band, we have shown that medium ($\approx10^2$) to high ($\approx10^3$) contrast detections can realistically be achieved for separations down to half of $\lambda/D$ on ultracool brown dwarf primary targets. In practice, this means that a 80-minute observation sequence can allow for the detection of a $1M_J$ situated $1.5\;AU$ away from a $30M_J$ Y type brown dwarf at a distance of $8$ pc. On brighter targets, kernel-phase analysis, combined with the methods presented in this paper can reveal companions at contrasts $\approx10^3$ down to $ 0.3 \lambda/D$.

Detection results presented in this paper rely on up-to-date simulations of JWST-NIRISS frames, that take into account all the noises expected to contribute to kernel-phase uncertainties. These results  can be affected by several effects that  are not yet accounted for, the most critical being probably  calibration errors. Instrumental drifts in the range of a few tens of nanometres, as predicted by \citet{Perrin2018UpdatedStrategies}, are not expected to degrade performances significantly for Y dwarfs. Another limitation may come from the algorithmic efficiency in  determining the MLE $\widehat{\boldsymbol{x}}$ in Eq. \eqref{eq:MLE} for the test $T_B$. Too coarse grid searches, or algorithms too sensitive to local minima will lead to a loss in detection power and to an increased uncertainty for the estimated parameters.

The performance reported in this work can therefore be seen as ideal contrast performance achievable using kernel phases for JWST NIRISS images. The method can in principle be improved upon by exploiting the full information available in the image (present not only in the phase but also, to a lesser extent, in the amplitude of the complex visibility). Even working solely with the phase, the calibration problem can be mitigated by using a more accurate and less idealised representation of the instrument. A significant fraction of the calibration error comes from the use of a necessarily approximate discrete model to represent the continuous phenomenon of diffraction. The results reported in this work used a dense aperture model to mitigate this discretisation error however the representation is not optimal yet. One avenue to improve the overall fidelity for example seems to be to take into account a variable local transmission function to more accurately describe the aperture with the same grid density. The study of the general aperture modelling prescription will be the object of future work.

The \texttt{XARA} package is regularly updated in the context of the KERNEL project.
\begin{acknowledgements}
KERNEL has received funding from the European Research Council (ERC) under the European Union's Horizon 2020 research and innovation program (grant agreement CoG - \#683029). We wish to thank Nick Cvetojevic for proof-reading the paper and his contribution to group discussions. The manuscript was also significantly improved following the comments and recommendations from the anonymous referee.
\end{acknowledgements}
\bibliographystyle{aa}
\inputencoding{latin2}
\bibliography{Main_text.bbl}
\inputencoding{utf8}
\clearpage

\end{document}